\documentclass[aps,pra,superscriptaddress]{revtex4-1}

\usepackage{preamble}
\usepackage[utf8]{inputenc}
\DeclareUnicodeCharacter{0338}{\o}
\graphicspath{{figs/}}
\newcommand{\scD}{\mathscr{D}}

\begin{document}
\title{Multiphase Porous Electrode Theory}

\author{Raymond B. Smith}
\affiliation{Department of Chemical Engineering, Massachusetts
Institute of Technology, Cambridge, Massachusetts 02139, USA}

\author{Martin Z. Bazant}
\email[Corresponding author: ]{bazant@mit.edu}
\affiliation{Department of Chemical Engineering, Massachusetts
Institute of Technology, Cambridge, Massachusetts 02139, USA}
\affiliation{Department of Mathematics, Massachusetts Institute of
Technology, Cambridge, Massachusetts 02139, USA}
\date{\today}

\begin{abstract}
Porous electrode theory, pioneered by John Newman and collaborators, provides a useful macroscopic description of battery cycling behavior, rooted in microscopic physical models rather than empirical circuit approximations.
The theory relies on a separation of length scales to describe transport in the electrode coupled to intercalation within small active material particles.
Typically, the active materials are described as solid solution particles with transport and surface reactions driven by concentration fields, and the thermodynamics are incorporated through fitting of the open circuit potential.
This approach has fundamental limitations, however, and does not apply to phase-separating materials, for which the voltage is an emergent property of inhomogeneous concentration profiles, even in equilibrium.
Here, we present a general theoretical framework for ``multiphase porous electrode theory'' implemented in an open-source software package called ``MPET'', based on electrochemical nonequilibrium thermodynamics.
Cahn-Hilliard-type phase field models are used to describe the solid active materials with suitably  generalized models of interfacial reaction kinetics.
Classical concentrated solution theory is implemented for the electrolyte phase, and Newman's porous electrode theory is recovered in the limit of solid-solution active materials with Butler-Volmer kinetics.
More general, quantum-mechanical models of Faradaic reactions are also included, such as Marcus-Hush-Chidsey kinetics for electron transfer at metal electrodes, extended for concentrated solutions.
The full equations and numerical algorithms are described, and a variety of example calculations are presented to illustrate the novel features of the software compared to existing battery models.
\end{abstract}

\maketitle

\section{Introduction}
Lithium-based batteries have growing importance in global society~\cite{scrosati2010} as a result of increased prevalence of portable electronic devices~\cite{tarascon2001}, and their enabling role in the transition toward renewable energy sources~\cite{whittingham2014}.
For example, lithium batteries can help mitigate intermittency of renewable energy sources such as solar power, and lithium battery powered electric vehicles are facilitating movement away from liquid fossil fuels for transportation.
Each of these growing areas demands high performance batteries, with requirements specific to the particular needs of the application driving specialized battery design for sub-markets.
Thus, it is critical that battery models be based on the underlying physics, enabling them to greatly facilitate cell design to take best advantage of the existing battery technologies.

\begin{figure}[h]
    \centering
    \includegraphics[width=0.4\textwidth]{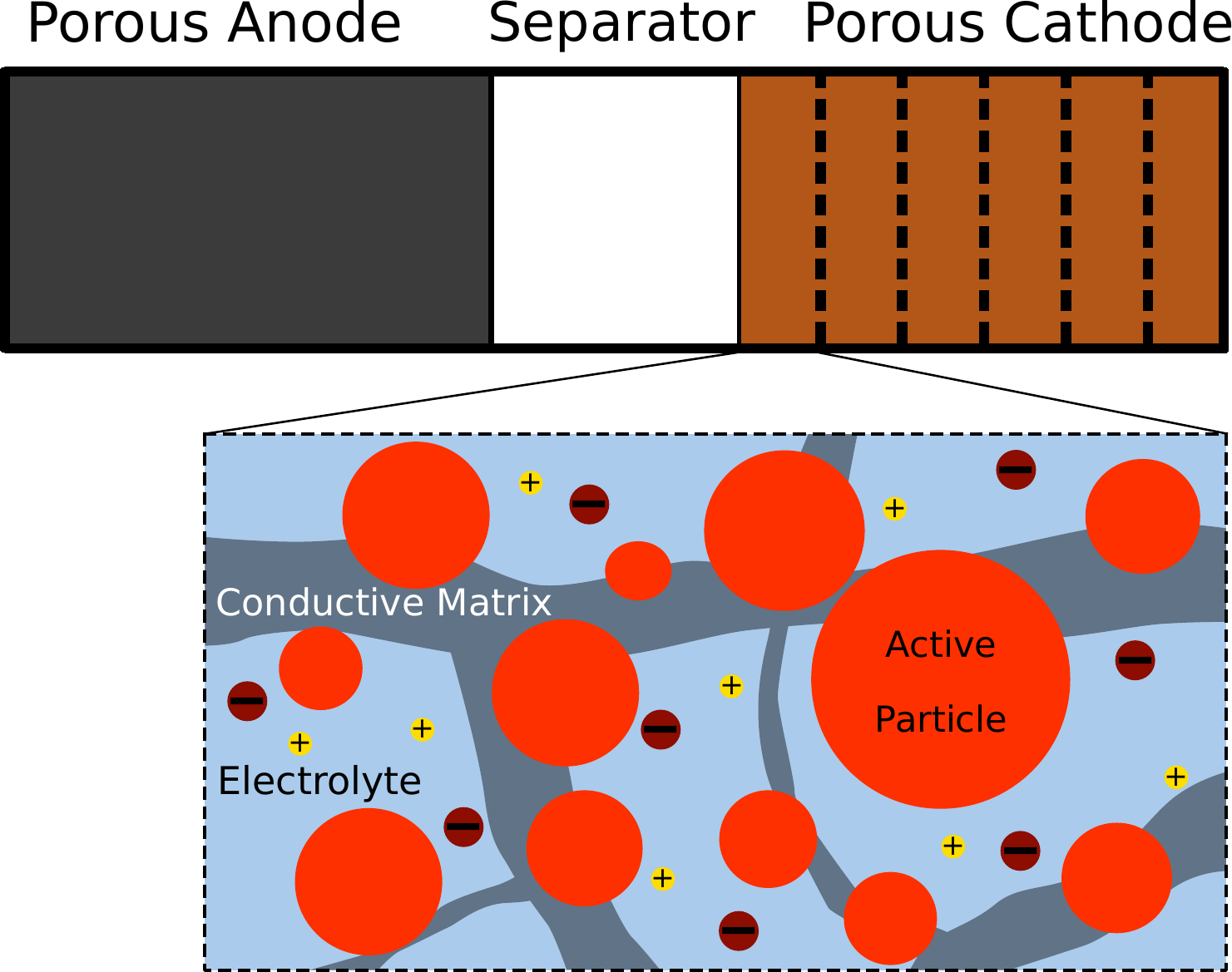}
    \caption{Schematic of a simulation of a battery with porous electrodes. The electrode is divided into finite volumes, and in each volume, a number of particles of active material interact with common electrolyte.}
    \label{fig:cell_schematic}
\end{figure}

Lithium-ion batteries are generally constructed using two porous electrodes and a porous separator between them.
The porous electrodes consist of various interpenetrating phases including electrolyte, active material, binder, and conductive additive.
A schematic is shown in Figure~\ref{fig:cell_schematic}.
In a charged state, most of the lithium in the cell is contained in the active material within the negative electrode.
During discharge, the lithium undergoes transport to the surface of the active material, electrochemical reaction to move from the active material to the electrolyte, transport through the electrolyte to the positive electrode, and reaction and transport to move into the active material of the positive electrode~\cite{winter2004,newman2004}.
Physical models must capture each of these behaviors accurately.

Complicating the situation further, the microstructure of the interpenetrating porous media within the electrodes can have a strong effect on the cell behavior~\cite{smith2009} as a result of inhomogeneities over a length scale smaller than the electrode but many times the size of a primary active particle~\cite{garcia2005,kehrwald2011,novak2000advanced,ebner2014,harris2013}.
The active materials themselves also often have highly non-trivial behavior including poor connectivity with each other and the conductive matrix~\cite{stephenson2007modeling,gaberscek2007wiring,orvananos2014architecture,ding2014dual} as well as complex material properties leading to deformations and accompanying stresses~\cite{aziz1991,christensen2006,tang2009,cheng2010diffusion,bower2011finite,dileo2015,woodford2010,sethuraman2012realtime,qi2010insitu,cogswell2012,ulvestad2014single,cannarella2014stress} and phase separation~\cite{whittingham2004,reimers1992,padhi1997,chen2006,ariyoshi2004,ohzuku1995,dahn1991,singer2014} during the intercalation process.
Coupling these behaviors in multi-particle electrode environments leads to further complexities~\cite{christensen2006,dreyer2010,dargaville2010predicting,li2014,zhang2014rate,ferguson2012,dargaville2013comparison,bai2013,levi2013collective,ferguson2014,orvananos2015effect,orvananos2014particle}.

Because capturing all the relevant physical processes from the atomistic length scale to the cell pack level within a single simulation is computationally intractable, various approaches have been developed to simulate aspects of battery behavior~\cite{grazioli2016,miranda2015}.
In particular, porous electrode theory, pioneered by John Newman and co-workers over the past fifty years, has proven highly successful in describing the practical scale of individual cells~\cite{newman1975,newman1962,doyle1993,fuller1994,doyle1996,darling1997,arora1999,srinivasan2004,christensen2006,devidts1997,verbrugge2003,zhang2007moving,dees2005,dees2009,baker2012,bernardi2011,gallagher2012,lai2011}.
This approach is based on volume averaging over a region of the porous electrode large enough to treat it as overlapping, homogeneous, continuous phases to describe the behavior of the electrons in the conductive matrix and the ions in the electrolyte.
The behavior of the active material is treated by defining representative particles and placing them within the simulation domain as volumetric source/sink terms for ions and electrons according to the actual volume fraction of active material in the electrode.
In this way, details on the length scale of transport within small active material particles can be consistently coupled to volume averaged transport over much larger length scales.
Ref.~\cite{thomas2001measurement} provides an excellent overview of the fundamentals of the theory.

As a result of volume averaging, heterogeneities over intermediate length scales are lost, causing inaccuracies in predictions.
Efforts to capture these heterogeneities in simulations have had success in characterizing the consequences of the volume averaging procedure and providing more accurate alternatives, including refinements to microstructural parameters such as tortuosity in the volume averaged approach~\cite{garcia2005,garcia2007,bertei2013,ebner2014,latz2015thermal}.
Nevertheless, simulations including complete microstructure information are much more computationally expensive than volume averaged approaches, so the simpler approach retains value in situations requiring faster model calculation or development.

Porous electrode theory has been developed and tested for decades for a variety of battery materials, but strictly speaking, it can only describe solid-solution active materials, whose thermodynamics are uniquely defined by fitting the open circuit voltage versus state of charge, or \textit{average} composition.
Active materials with more complex thermodynamics, resulting in multiple stable phases of different equilibrium concentrations, cannot be described, except by certain empirical modifications.
Phase separating materials, such as lithium iron phosphate and graphite, can be accommodated by introducing artificial phase boundaries, such as shrinking cores~\cite{srinivasan2004} or shrinking annuli~\cite{hess2013}, respectively, but this approach masks the true thermodynamic behavior.

Instead, the open circuit voltage of a battery is an emergent property of multiphase materials, which reflects phase separation in single particles~\cite{bai2011,cogswell2012,cogswell2013,deklerk2016anatase_draft} and porous electrodes~\cite{lai2011,ferguson2012,ferguson2014}.
It can only be predicted by modeling the free energy functional, rather than the voltage directly, and consistently defining electrochemical activities, overpotentials, and reaction rates using variational nonequilibrium thermodynamics~\cite{bazant2013,bazant2017thermodynamic}.
This is the approach of ``multiphase porous electrode theory'' presented below.
In principle, such models are required to predict multiphase battery performance over a wide range of temperatures and currents~\cite{ferguson2012,ferguson2014,smith2017_intercalationdraft,bazant2017thermodynamic}, as well as degradation related to mechanical stresses~\cite{cogswell2012,cogswell2013} and side reactions that depend on the local surface concentration profile~\cite{thomas-alyea2017_submitted}.

Regardless of the thermodynamic model, volume averaged simulations using porous electrode theory are carried out in a number of ways.
Newman's \texttt{dualfoil} code uses a finite difference method via the BAND subroutine, and it is freely available~\cite{newmandualfoil} and commonly used.
It has been developed and tested for decades, but it requires analytical derivative information about model equations to form the Jacobian, which makes modifications to the code less straightforward.
The free energy approach we take here naturally describes both phase separating and solid solution materials using the same mathematical framework in a more user-friendly implementation.
Popular commercial software packages such as COMSOL~\cite{comsol} have also been used to implement versions of porous electrode theory~\cite{stephenson2007modeling,bernardi2011,dees2009,cai2011mathematical}, usually using the finite element method.
This has the advantage of being quick to set up, but it can involve costly software licenses.
More importantly, the closed source means that detailed inspection of the software is impossible for the purpose of verifying, modifying, and improving the numerical methods for the particular problems investigated.
A number of authors have also written custom versions of porous electrode software~\cite{ferguson2012,dargaville2010predicting,torchio2016lionsimba}, for example using a manually implemented finite volume method and some general differential algebraic equation (DAE) solver for time advancement.
This approach provides significant flexibility, but it is not common to share the code to facilitate use and inspection by a broader community, although Torchio et al.\ recently published their MATLAB implementation using the finite volume method with a variable time stepper~\cite{torchio2016lionsimba}.
We take a similar approach in this work.
More comprehensive reviews of commonly used simulation approaches can be found in refs.~\cite{botte2000mathematical,ramadesigan2012}.

Here, we present the equations and algorithms for a finite-volume based simulation software package, which implements multiphase porous electrode theory (``MPET'').
The code is freely available~\cite{smithmpet}, and it is developed in a modular way to facilitate modification and re-use.
It is based only on open-source software and is written in Python, a modern, high-level language commonly used in the scientific computing community.
Computationally expensive aspects of the code are all done using standard and freely available, open source numerical libraries.
This takes advantage of Python's ease of use for the model definitions while retaining the fast and vetted computation of libraries written in lower level languages like C and Fortran.
%We present here the basic framework of the model and implementation in which we extend the framework developed by Ferguson and Bazant~\cite{ferguson2012}.
%First, we add the option to simulate Stefan-Maxwell, concentrated electrolyte solutions necessary to describe cells with substantial polarization~\cite{newman2004,valoen2005}.
%Second, for the active material models, we capture the continued development of models pursued by Bazant and co-workers describing a wide range of physical models for the bulk solid~\cite{singh2008,burch2009,bai2011,ferguson2014,smith2017_intercalationdraft}, many of which have been experimentally validated~\cite{ferguson2014,li2014,lim2016origin,guo2016,thomas-alyea2017_submitted,deklerk2016anatase_draft}.
%Finally, we generalize it to be able to easily simulate single particles, half-cells with a Li-foil counter-electrode, and full porous electrodes.
%We also add and highlight here the ability to study different models for electron transfer reaction kinetics to facilitate reexamination of the use of Butler-Volmer~\cite{bard2001,newman2004,dreyer2016,heubner2015} or Marcus~\cite{marcus1956oxidationreduction,marcus1965electron,marcus1993,chidsey1991,bai2014} style kinetic expressions in porous electrode battery modeling.

The paper is organized as follows.
In Section~\ref{sec:model}, we begin by presenting the full mathematical framework of MPET, based on the original formulation of Ferguson and Bazant~\cite{ferguson2012}, with several modifications.
First, we incorporate the standard description of transport in concentrated electrolytes based on Stefan-Maxwell coupled fluxes and chemical diffusivities~\cite{newman2004,valoen2005}.
Second, we capture the continued development by our group of phase field models for the active solid materials~\cite{bazant2013,singh2008,burch2009,bai2011,cogswell2012,cogswell2013,ferguson2014,smith2017_intercalationdraft,deklerk2016anatase_draft}, which have increasingly been validated by direct experimental observations of phase separation dynamics~\cite{ferguson2014,li2014,lim2016origin,bazant2017thermodynamic,guo2016,thomas-alyea2017_submitted,deklerk2016anatase_draft}.
Third, we provide alternatives to the empirical Butler-Volmer model of Faradaic reaction kinetics~\cite{bard2001,newman2004,dreyer2016,heubner2015}, based on the quantum mechanical theory of electron transfer pioneered by Marcus~\cite{marcus1956oxidationreduction,marcus1993,marcus1965electron,chidsey1991} and extended here for concentrated solutions~\cite{smith2017mhc_draft}, motivated by recent battery experiments~\cite{bai2014}.
Fourth, the porous electrode model is modified to allow for different network connections between active particles, as well as half-cells with Li-foil counter electrodes and full two-electrode cells.
In Section~\ref{sec:ndim_eqns}, the equations are made dimensionless, and numerical methods to solve them are presented in Section~\ref{sec:implementation}, along with the overall software structure.
In Section~\ref{sec:examples}, a variety of example simulations are presented to highlight the novel features of MPET compared to previous models, and the paper concludes with an outlook for future developments in Section~\ref{sec:concl}.

\section{Model}
\label{sec:model}
As discussed above, the basic structure of the model involves volume averaging over a region larger than the particles of active material.
Because the details of dynamics within the active material particles can strongly affect model predictions, they are simulated at their small length scale and treated as a source term for equations defined over the larger, electrode length scales.
Thus, the model can be broken down into a number of scales.
For the overall cell, we specify either the cell current density or voltage input profiles along with any series resistance.
The unspecified current or voltage is an output of the simulation.
At the electrode scale, we solve electrolyte transport equations, potential losses in the electron-conducting matrix, and potential drop between particles.
At the particle scale, we simulate how they react with the electrolyte and the internal concentration dynamics.
We will assume uniform (though not necessarily constant) temperature in the model derivation. The software currently only supports uniform and constant room temperature simulations, but we retain temperature factors here for generality.
We will use the convention in referring to electrodes that the electrode which is negative/positive at open circuit and charged conditions is referred to as the anode/cathode.

\subsection{Electrode Scale Equations}
\subsubsection{Electrolyte Model}
\label{sec:model_elyte}
The general form of the electrolyte model equations arises from statements of conservation of species and conservation of charge within the electrolyte phase of a quasi-neutral porous medium~\cite{newman1962,devidts1997}.
We consider electrolytes of salts defined by $\sum_{i}\nu_{i}M_{i}^{z_i}$, with species $M_i$ having valence $z_i$, and $\nu_i$ the number of ions $M_i$ in solution from dissolving one molecule of the neutral salt.
For example, for CuCl$_2$, $\nu_+ = 1$, $M_+ = \mathrm{Cu}$, $z_+ = 2$, $\nu_- = 2$, $M_- = \mathrm{Cl}$, and $z_- = -1$.
With electrolyte species flux, $\mathbf{F}_{\ell,i}$ defined per area of porous medium, conservation of species requires
\begin{align}
    \frac{\partial\left( \eps c_{\ell,i} \right)}{\partial t} = -\dvg\mathbf{F}_{\ell,i} + R_{V,i}
    \label{eq:cConsFlux}
\end{align}
where $c_{\ell,i}$ is the concentration of species $i$ in the electrolyte, $\eps$ is the electrolyte volume fraction (porosity), and $R_{V,i}$ describes a volume-averaged reaction rate which is the result of interfacial electrochemical reactions with the active materials in which ions are added to or removed from the electrolyte.

In general, electrolyte transport models for porous media, the macroscopic charge density may be nonzero and varies in response to current imbalances.
Diffuse electrolyte charge screens internal charged surfaces of porous membrane materials~\cite{dydek2011,dydek2013,schmuck2015homogenization,yaroshchuk2012over} or conducting porous electrodes~\cite{biesheuvel2011diffuse,mirzadeh2014enhanced} and provides additional pathways for ion transport by electromigration (``surface conduction'') and electro-osmotic flows.
In addition to Faradaic reactions, porous electrodes can also undergo capacitive charging by purely electrostatic forces, as in electric double layer capacitors and capacitive deionization systems~\cite{biesheuvel2011diffuse}, or hybrid pseudo-capacitors~\cite{biesheuvel2012electrochemistry}.
In batteries, however, such effects are usually neglected~\cite{newman1975,thomas2002,newman2004}, since the focus is on electrochemical, rather than electrostatic, energy storage, using highly concentrated electrolytes.

In these electrolyte models, we will assume they satisfy quasi-neutrality, i.e.\@ there is no net charge in the electrolyte over the simulated length scales~\cite{dickinson2011}.
Charge conservation and quasi-neutrality together require
\begin{align}
    \frac{\partial\left( \eps\rho_e \right)}{\partial t} \approx 0 = -\dvg\mathbf{i}_{\ell}
    + \sum_i z_{i}eR_{V,i},
    \label{eq:qConsFlux}
\end{align}
where $e$ is the elementary charge, the charge density $\rho_e = \sum_i z_{i}ec_{\ell,i}$, and $\mathbf{i}_{\ell}$ is the current density in the electrolyte, related to a sum of ionic fluxes,
\begin{align}
    \mathbf{i}_{\ell} &= \sum_i z_{i}e\mathbf{F}_{\ell,i}.
%    \\
%    &= -D_{\mathrm{chem},+}
    \label{eq:currDensDef}
\end{align}
We will relate fluxes to both concentrations and electrostatic potentials in the electrolyte, so with constitutive flux relationships, Eqs.~\ref{eq:cConsFlux} and~\ref{eq:qConsFlux} fully specify the system for both the set of concentrations and the electrostatic potential field.
However, it is convenient to use quasi-neutrality to eliminate one of the species conservation equations using (with arbitrary $n$)
\begin{align}
    z_{n}c_{\ell,n} = -\sum_{i\ne n}z_{i}c_{\ell,i},
    \label{}
\end{align}
which allows us to neglect Eq.~\ref{eq:cConsFlux} for one species and post-calculate the missing concentration profile.
For example, for the case of a binary electrolyte of cations, $+$, and anions, $-$, we define the neutral salt concentration, $c_\ell = \frac{c_{\ell,+}}{\nu_+} = \frac{c_{\ell,-}}{\nu_-}$ and instead simulate one of
\begin{align}
    \frac{\partial\left( \eps c_\ell \right)}{\partial t} =
%    \dvg\left(\eps\left( D_{\mathrm{chem},i}\bnab c + D_{i}z_{i}c\bnab\wt{\phi} \right)\right)
    \frac{1}{\nu_i}\Big(-\dvg\mathbf{F}_{\ell,i} + R_{V,i}\Big).
    \label{eq:cSaltConsFlux}
\end{align}
and
\begin{align}
    0 = -\dvg\mathbf{i}_{\ell} + \left( z_{+}eR_{V,+} + z_{-}eR_{V,-} \right).
    \label{eq:qConsFluxBinary}
\end{align}
In the case of Li-ion batteries for which we will assume a binary electrolyte in which only the Li$^+$ ions react, $R_{V,-} = 0$, so it is particularly convenient to simulate the anion species conservation equation.

The boundary conditions for a porous electrode simulation relate the fluxes of the simulated species and current at the anode and cathode current collectors
and depend on the simulation.
%At both current collectors in a Li-ion battery, $\wh{\mathbf{n}}\cdot\mathbf{N}_{\ell,i} = 0$.
At the current collectors of porous electrodes, $\wh{\mathbf{n}}\cdot\mathbf{F}_{\ell,i} = 0$ and $\wh{\mathbf{n}}\cdot\mathbf{i}_{\ell} = 0$.
However, if a foil (e.g.\ Li metal electrode) is used for one electrode, the boundary condition at that side is replaced by $\wh{\mathbf{n}}\cdot\mathbf{i}_{\ell} = i_\mathrm{cell}$, where $i_\mathrm{cell}$ is the macroscopic current density of the cell, Eq.~\ref{eq:Icell}. The species flux boundary conditions depend on which species is being eliminated in the set of species conservation equations, but for the case of anion conservation in a Li-ion battery, $\wh{\mathbf{n}}\cdot\mathbf{F}_{\ell,-} = 0$ at all current collectors (neglecting side reactions).

The transport in the electrolyte can be simulated using either simple dilute Nernst-Planck equations or a concentrated solution model based on Stefan-Maxwell coupled fluxes.
In both cases, we will neglect convection and assume binary electrolytes.
Both formulations relate a species flux, $\mathbf{F}_{\ell,i}$ to gradients in electrochemical potentials, $\mu_{\ell,i}$.
The electrochemical potential of a given ion generally has both chemical and electrostatic contributions, and these can be separated a number of ways.
For example, it can be separated using an ``inner'' or Galvani potential~\cite{vetter1967,newman2004},
\begin{align}
    \mu_{\ell,i} = k_\mathrm{B}T\ln\left( a_{\ell,i} \right) + \mu_{\ell,i}^\Theta + z_{i}e\phi_{\ell} = k_\mathrm{B}T\ln\left( \wt{c}_{\ell,i} \right) + \mu_{\ell,i}^\mathrm{ex}
    \label{eq:mu_galvani}
\end{align}
where $k_\mathrm{B}$ is the Boltzmann constant, $T$ is the absolute temperature, $a_{\ell,i}$ is the activity of species $i$, $\mu_{\ell,i}^\Theta$ is its reference chemical potential, and $\phi_{\ell}$ is the (Galvani) electrostatic potential.
The final expression serves as the definition of the excess chemical potential, $\mu_{\ell,i}^\mathrm{ex}$, and $\wt{c}_{\ell,i}$ is the concentration, $c_{\ell,i}$, scaled to some suitable reference.
The excess chemical potential contains all of the entropic and enthalpic contributions to the free energy of a concentrated solution beyond that of a dilute solution of a neutral species, such as short-ranged forces, long-ranged electrostatics, and excluded volume effects for finite sized ions~\cite{bazant2009towards,liu2014solute}.
%It determines the equilibrium phase diagram of the electrolyte.

An alternative approach common in battery modeling~\cite{doyle1993,thomas2002,verbrugge2003} for separating these contributions is to define the electrostatic potential in the electrolyte as that measured with a suitable reference electrode at a position of interest in the electrolyte with respect to another reference at a fixed position in the solution.
This has the advantage, unlike Eq.~\ref{eq:mu_galvani}, of being defined entirely by measurable quantities, unlike the activities of individual ions in solution.
For example, in a lithium ion battery, this is typically done with a Li/Li$^+$ reference electrode.
We will refer to this potential as $\phi_{\ell}^r$, and we note that it is actually measuring (a combination of) full electrochemical potentials of ions in solution.
For the Li/Li$^+$ example,
\begin{align}
    e\phi_{\ell}^r = \mu_{\ell,\mathrm{Li}^+} + \cnst = k_\mathrm{B}T\ln\left( a_{\ell,\mathrm{Li}^+} \right) + e\phi_{\ell} + \mu_{\ell,\mathrm{Li}^+}^\Theta + \cnst.
    \label{}
\end{align}
Quasi-electrostatic potentials, $\phi_{\ell}^q$, can also be used in electrolyte models~\cite{newman2004}, and they are defined by the excess chemical potential of a particular ion in solution, $n$,
\begin{align}
    z_{n}e\phi_{\ell}^q = \mu_{\ell,n}^\mathrm{ex} = z_{n}e\phi_{\ell} + k_\mathrm{B}T\ln\left( \gamma_{\ell,n} \right) + \mu_{\ell,n}^\Theta
    \label{}
\end{align}
where $\gamma_{\ell,i} = \frac{a_{\ell,i}}{\wt{c}_{\ell,i}}$ is the activity coefficient of species $i$.

We also note here that the electric potential field is determined in electrolytes either (1) by an assumption of quasi-neutrality, i.e.\ $\sum_i z_{i}c_{\ell,i} = 0$ everywhere, or (2) by solving the Poisson equation, $\dvg\left( \varepsilon\bnab\phi_{\ell} \right) = -\rho_e$ with permittivity $\varepsilon$, which could depend on concentration or electric field~\cite{bazant2009towards}, or capture non-local ion-ion correlations as a differential operator~\cite{bazant2011double}.
Throughout this work, we assume quasi-neutrality but make a few comments about the alternative here.
Use of the Poisson equation enables physical boundary conditions on the electric potential such as specified surface charge densities at interfaces, and the resulting electric potential is the potential of mean force acting on a test charge in the solution.
It captures double-layers of diffuse charge with thickness characterized by the Debye length at interfaces, outside of which the net charge approaches zero.
The assumption of quasi-neutrality leads to a different potential field which cannot capture the effects of electric double layers at interfaces.
Of note, the form of the chemical potential which can most easily accommodate models with both the quasi-neutral and the Poisson equations is the Galvani potential.
Of course, this can lead to further confusion because the potential field obtained by assuming quasi-neutrality need not satisfy the Poisson equation with zero charge density and generally will not; rather, quasi-neutrality in this case is the result of the ``outer'' solution to a singular perturbation in which the Debye length approaches zero.
Thus, we could further distinguish between Galvani potentials obtained via each of those methods, but refrain from complicating the notation here, as the quasi-neutral Galvani potential closely resembles the Poisson-satisfying Galvani potential in the large majority of systems with dimensions much larger than the Debye length.

\paragraph{(Semi-)Dilute Electrolyte}
In the simpler model, we neglect couplings between species fluxes, such that the flux of a given ionic species is a function only of gradients in its own electrochemical potential,
\begin{align}
    \mathbf{F}_{\ell,i} = -M_{\ell,i}c_{\ell,i}\bnab\mu_{\ell,i}
    \label{}
\end{align}
where $M_{\ell,i}$ is the species' mobility.
Using the Galvani potential form in Eq.~\ref{eq:mu_galvani},
\begin{align}
    \mathbf{F}_{\ell,i} &= -\frac{D_{\ell,i}}{\wt{T}}\left(c_{\ell,i}\bnab \left(\wt{T}\ln\left( a_{\ell,i} \right)\right) + z_{i}c_{\ell,i}\bnab\wt{\phi}_{\ell}\right),
    \label{}
\end{align}
where we have used the Einstein relation between the diffusivity in free solution, $D_{\ell,i}$ and mobility, $D_{\ell,i} = M_{\ell,i}k_\mathrm{B}T$, non-dimensionalized the potential by the thermal voltage at some reference temperature, $\wt{\phi}_{\ell} = \frac{e\phi_{\ell}}{k_\mathrm{B}T_\mathrm{ref}}$, and defined a non-dimensional $\wt{T} = T/T_\mathrm{ref}$.
With $D_{\ell,\mathrm{chem},i} = D_{\ell,i}\left( 1 + \frac{\partial\ln\gamma_{\ell,i}}{\partial\ln\wt{c}_{\ell,i}} \right)$ and uniform temperature,
%and $\gamma_i = \exp{\left(\left( \mu_i^\mathrm{ex} - \mu_i^\Theta \right)/\left(k_\mathrm{B}T\right)\right)}$,
%Further assuming the solution is dilute such that we can use $a_i \propto \wt{c}_i$, then
\begin{align}
    \mathbf{F}_{\ell,i} = -\left(D_{\ell,\mathrm{chem},i}\bnab c_{\ell,i} - \frac{D_{\ell,i}}{\wt{T}}z_{i}c_{\ell,i}\bnab\wt{\phi}_{\ell}\right).
    \label{}
\end{align}
In a porous medium we use effective transport properties, which are adjusted by the tortuosity of the electrolyte phase, $\tau$.
In addition we define the flux per area of porous medium rather than per area of electrolyte requiring a prefactor of the porosity of the electrolyte phase, $\eps$.
\begin{align}
    \mathbf{F}_{\ell,i} &= -\frac{\eps}{\tau}\left(D_{\ell,\mathrm{chem},i}\bnab c_{\ell,i} + \frac{D_{\ell,i}}{\wt{T}}z_{i}c_{\ell,i}\bnab\wt{\phi}_{\ell}\right).
    \label{eq:NernstPlanckDchem}
\end{align}
The tortuosity is often described as a function of the porosity, the volume fraction of the electrolyte phase, $\eps$, commonly by employing the Bruggeman relation $\tau = \eps^{a}$~\cite{bruggeman1935}.
The value of $a$ is often set to $-0.5$ but can be adjusted~\cite{shen2007} to account for experimentally~\cite{thorat2009quantifying,shearing2010characterization,wilson2011measurement,ender2011three,ebner2014} or theoretically~\cite{vijayaraghavan2012,torquato2013,garcia-garcia2016} observed departures from the original derivation.
Eq.~\ref{eq:NernstPlanckDchem} with Eqs.~\ref{eq:currDensDef},~\ref{eq:cSaltConsFlux} and~\ref{eq:qConsFluxBinary} define the (semi-)dilute electrolyte model.
Although the electrolyte transport model developed in the following section is more reasonable for battery models, we present and retain the (semi-)dilute model here for a number of reasons.
Retaining it facilitates comparisons between the models, and the dilute model is easier to extend for electrolytes with more components, even if doing so loses the information related to the extra transport parameters associated with Stefan-Maxwell transport theories.
In addition, as mentioned above, it is straightforward to connect the Galvani potential used here to extensions using the Poisson equation to investigate behaviors at interfaces.

\paragraph{Stefan-Maxwell Concentrated Electrolyte}
The formulation above assumes that gradients in the electrochemical potential of species $i$ lead to fluxes only of species $i$.
However, the framework can be generalized by assuming that fluxes of a given species are related to gradients in electrochemical potentials of each species in the system,
\begin{align}
    \mathbf{F}_{\ell,i} = \sum_{j}U_{ij}\bnab\mu_{\ell,j}.
    \label{}
\end{align}
where $U_{ij}$ are the direct ($i=j)$ and indirect ($i \ne j$) transport coefficients~\cite{balluffi1954}.
This formulation has been used to describe relatively concentrated electrolytes and is the most commonly used model in battery simulation~\cite{newman2004,thomas2002}.
Noting that not all electrochemical potentials are independent (from the Gibbs-Duhem relationship), the above can be reorganized to the more commonly written form in terms of species velocities, $\mathbf{v}_i = \mathbf{F}_{\ell,i}/c_{\ell,i}$,~\cite{newman2004}
\begin{align}
    c_{\ell,i}\bnab\mu_{\ell,i} = \sum_{j}K_{ij}\left( \mathbf{v}_j - \mathbf{v}_i \right)
    = k_\mathrm{B}T\sum_{j}\frac{c_{\ell,i}c_{\ell,j}}{c_{T}\mathscr{D}_{ij}}\left( \mathbf{v}_j - \mathbf{v}_i \right)
    \label{}
\end{align}
with $c_{T} = \sum_{i}c_{\ell,i}$ and $K_{ij} = K_{ji}$.
%and $R = N_\mathrm{A}k_\mathrm{B}$ is the gas constant and $N_\mathrm{A}$ is the Avogadro constant.
For a binary electrolyte in a porous medium with cations, anions, and solvent (denoted as species $0$), assuming uniform temperature and that the solvent concentration varies only negligibly with salt concentration, and again neglecting convection,
\begin{align}
    \mathbf{F}_{\ell,+} &= -\frac{\nu_{+}\eps}{\tau}D_\ell\bnab c_{\ell} + \frac{t_{+}^{0}\mathbf{i}_{\ell}}{z_{+}e}
    \label{eq:SM_NpD}
    \\
    \mathbf{F}_{\ell,-} &=  -\frac{\nu_{-}\eps}{\tau}D_\ell\bnab c_{\ell} + \frac{t_{-}^{0}\mathbf{i}_{\ell}}{z_{-}e}
    \label{eq:SM_NmD}
\end{align}
where, defining $\gamma_{\ell,\pm}^\nu = \gamma_{\ell,+}^{\nu_+}\gamma_{\ell,-}^{\nu_-}$ with $\nu=\nu_+ + \nu_-$,
\begin{align}
    D_{\ell} &= \mathscr{D}\frac{c_T}{c_{\ell,0}}\left( 1 + \frac{\partial\ln \gamma_{\ell,\pm}}{\partial\ln\wt{c}_{\ell}} \right),
    \\
    \scD &= \frac{\scD_{0+}\scD_{0-}\left( z_+ - z_- \right)}{z_+\scD_{0+} - z_-\scD_{0-}},
    \\
    t_+^0 &= 1 - t_-^0 = \frac{z_+\scD_{0+}}{z_+\scD_{0+} - z_-\scD_{0-}}.
    \label{}
\end{align}
The current density is given by
\begin{align}
%    \mathbf{i} = -\frac{\kappa\eps}{\tau}\left(\bnab\phi^r
    \mathbf{i}_{\ell} = -\frac{\sigma_{\ell}\eps}{\wt{T}\tau}\left(\bnab\phi_{\ell}^r
    + \frac{\nu k_\mathrm{B}T}{e}
    \left( \frac{s_+}{n\nu_+} + \frac{t_+^0}{z_{+}\nu_{+}} - \frac{s_{0}c_{\ell}}{nc_{\ell,0}} \right)
    \left( 1 + \frac{\partial\ln \gamma_{\ell,\pm}}{\partial\ln \wt{c}_{\ell}} \right)\bnab\ln \wt{c}_{\ell}\right)
    \label{eq:currDens}
\end{align}
with
\begin{align}
%    \frac{1}{\kappa} &=
    \frac{1}{\sigma_{\ell}} &=
    -\left( \frac{k_\mathrm{B}T_\mathrm{ref}}{c_{T}z_{+}z_{-}e^2} \right)
    \left( \frac{1}{\scD_{+-}}
    + \frac{c_{\ell,0}t_{-}^0}{c_{\ell,+}\scD_{0-}} \right).
    \label{eq:kappa_defn}
\end{align}
The values of $s_i$ are specified by the choice for the reference electrode with reaction
\begin{align}
    s_{-}M_{-}^{z_-} + s_{+}M_{+}^{z_+} + s_{0}M_{0} \rightleftharpoons n\mathrm{e}^{-}.
    \label{}
\end{align}
For lithium-ion batteries, the typical choice for the reference electrode defining $\phi_{\ell}^r$ is Li/Li$^+$, so $s_+ = -1$, and $s_- = s_0 = 0$.
Thus, Eqs.~\ref{eq:SM_NmD} and~\ref{eq:currDens} with Eqs.~\ref{eq:cSaltConsFlux} and~\ref{eq:qConsFluxBinary} define the electrolyte model when using the Stefan-Maxwell concentrated solution theory with a binary electrolyte.

\subsubsection{Solid phase electronic model}
\label{sec:model_bulk_e_losses}
The solid phase of a porous electrode is composed of several length scales and percolating phases.
The active material stores the reduced species (e.g.\ lithium), conductive additive improves electronic wiring, and binder is added to keep all the components connected.
Lithium transport occurs within individual particles, and electrons must reach the surface of those particles via traveling over the length of the electrode.
These equations approximately capture the variation in electric potential in the matrix of conductive material, which may be assumed to be identical to that of (well connected) active material.
Otherwise, additional relations can be added to describe losses between the conductive matrix and the active materials.
Over the length scale of the electrode, we describe conservation of charge as in the electrolyte,
\begin{align}
    0 = -\dvg\mathbf{i}_s - \sum_{i}z_{i}eR_{V,i}
    \label{eq:consQsld}
\end{align}
or
\begin{align}
    0 = -\dvg\mathbf{i}_s - \left(z_{+}eR_{V,+} + z_{-}eR_{V,-}\right)
    \label{eq:consQsldBinary}
\end{align}
for a binary electrolyte in which cations and/or anions may undergo electrochemical reactions, and where $\mathbf{i}_s$ is the current density in the solid phase.
The sign difference compared to Eq.~\ref{eq:qConsFluxBinary} comes from the observation that charge entering the liquid phase must be leaving the solid phase.
The current density in the solid phase is given by assuming a bulk Ohm's law,
\begin{align}
    \mathbf{i}_s = -\frac{\left( 1-\eps \right)}{\tau}\sigma_s\bnab\phi_s
    \label{}
\end{align}
where $\sigma_s$ is the conductivity of the electronically conductive matrix, $\phi_s$ is its electrostatic potential, and we have assumed that the volume fraction of the conductive phase is given by the space not occupied by the electrolyte.
The boundary conditions for this come from observing no electronic current can flow into the separator, $\wh{\mathbf{n}}\cdot\mathbf{i}_s = 0$, and the potential at the current collector is specified by the operating voltage on the system in the macroscopic equations, $\phi_c$ or $\phi_a$ ($c$ stands for cathode while $a$ stands for anode).
The Bruggeman relation can again be used to estimate the tortuosity in terms of the volume fraction of the conductive matrix.

To avoid the computational and practical difficulties of simulating full microstructures, we describe the behavior of particle interactions with a small number of representative particles, both along the length of the electrode and also in parallel with each other in terms of electrolyte access to capture the effects of particle size distributions.
The particles at the same electrode position (interacting in common with the local electrolyte) could be in parallel or in series electronically where parallel wiring would describe them each having direct access via a single resistance to a conductive network and series wiring might describe a comb structure in which some particles (perhaps the edge of a secondary particle) are connected to the conductive backbone, but electrons must pass through poorly conducting particles to get to particles without good contact to the conductive backbone, similar in concept to the hierarchical model of Dargaville and Farrell~\cite{dargaville2010predicting}.
We demonstrate use of the parallel case with a distribution of contact resistances in ref.~\cite{thomas-alyea2017_submitted}.
In the second case of series wiring we implement a simplified version of that developed by Stephenson et al.~\cite{stephenson2007modeling} by imposing a finite conductance between particles in series, indexed by $k$, within a simulation volume, $j$,
\begin{align}
    G_{j,k}\left(\phi_{j,k} - \phi_{j,k+1}\right) = I_{j,k}
    \label{}
\end{align}
where $G_{j,k}$ is the conductance and $I_{j,k}$ is the current between particle $k$ and $k+1$.
From charge conservation,
\begin{align}
    I_{j,k} - I_{j,k+1} = \int_{S_{k+1}}^{}j_{j,k+1}\ud A,
    \label{}
\end{align}
where $j_{j,k+1}$ is the intercalation rate into particle $k+1$.

\subsection{Single Particle Equations}
A single particle interacts with the electrolyte via an electrochemical reaction, leading to intercalation of neutral species into the solid phase.
However the electrochemistry is modeled (see Section~\ref{sec:echemRxns}), the reaction serves as a source/removal of species into/from the particle.
We will describe several different solid models here.
Generally, we begin by postulating a free energy functional describing the important physics of the particle,
\begin{align}
%    G\left[\left\{c\left(\mathbf{r}\right)\right\}, \left\{\bnab c_{s,i}\right\}, \bs{\eps}\left( \mathbf{r} \right)\right]
    G
    = \int_{V_p}^{} g \ud V + \int_{A_S}^{}\gamma_S\ud A
    \label{}
\end{align}
where $G$ is the total system free energy, $V_p$ is the particle volume, $g$ is the free energy density, $A_S$ is the particle surface area, and $\gamma_S$ is the surface energy.
Typically, we will separate the free energy density into homogeneous, $g_\mathrm{h}$ and non-homogeneous, $g_\mathrm{nh}$, contributions,
\begin{align}
    g = g_\mathrm{h} + g_\mathrm{nh} + \dots
    \label{}
\end{align}
where the remaining terms could describe the stress state of the system~\cite{cogswell2012,bazant2013} or other energetic contributions~\cite{garcia2004}.
Following van der Waals~\cite{rowlinson1979translation} and Cahn and Hilliard~\cite{cahn1958}, we use a simple gradient penalty term to describe the non-homogeneous free energy,
\begin{align}
    g_\mathrm{nh} = \frac{1}{2}\frac{1}{c_{s,\mathrm{ref}}^2}\bnab c_{i} \cdot \bs{\kappa}\bnab c_{i}
    \label{}
\end{align}
where $\kappa$ is a gradient penalty tensor (assumed to be isotropic here such that $\bs{\kappa} = \kappa\mathbf{1}$ and $\mathbf{1}$ is the second-order identity tensor) related to interfacial energy between phases, $c_i$ is the concentration of species $i$ within a single particle, and $c_{s,\mathrm{ref}}$ is a suitable concentration scale for the insertion species in the active material.
The diffusional chemical potential can then be obtained form a variational derivative of the free energy,
\begin{align}
    \mu_i = \frac{\delta G}{\delta c_{i}} = \frac{\partial g}{\partial c_{i}} - \dvg\frac{\partial g}{\partial\bnab c_{i}}.
    \label{eq:muVarDef}
\end{align}

\subsubsection{Electrochemical Reactions}
\label{sec:echemRxns}
The electrochemical reaction can be described by a number of different models, such as the empirical Butler-Volmer equation~\cite{bard2001} or quantum-mechanical models based on Marcus kinetics~\cite{marcus1993,chidsey1991,zeng2014simple}, which must be consistently generalized for concentrated solutions in nonequilibrium thermodynamics~\cite{bazant2013,smith2017mhc_draft}.
We describe here electrochemical reactions of the form
\begin{align}
    S_1 = \sum_{i} s_{i,O}O_{i}^{z_{i,O}} + n\mathrm{e}^- \to
    \sum_{j} s_{j,R}R_{j}^{z_{j,R}} = S_2,
    \label{}
\end{align}
and we will describe the reactions as a function of the activation overpotential,
\begin{align}
    ne\eta = \mu_R - \left(\mu_O + n\mu_\mathrm{e}\right) = \Delta\mu_\mathrm{rxn} = \Delta G_\mathrm{rxn},
    \label{}
\end{align}
where $\mu_R = \sum_{j}s_{j,R}\mu_{j}$ is the electrochemical potential of the reduced state, $\mu_O = \sum_{i} s_{i,O}\mu_{i}$ is the electrochemical potential of the oxidized state, and $\mu_\mathrm{e}$ is the electrochemical potential of the electrons, which we relate here to the electric potential measured in the conductive matrix, $\mu_e = -e\phi_s$.
$\Delta\mu_\mathrm{rxn}$ and $\Delta G_\mathrm{rxn}$ indicate the total free energy change of the reaction.
In the case of Li$^+$ insertion and the Stefan-Maxwell concentrated electrolyte model using Li/Li$^+$ as a reference electrode, $\mu_O = e\phi_{\ell}^r$.
We adopt the convention here that a positive electrochemical reaction current corresponds to the net rate of reduction.
%In each model, we will describe the net reduction current, $i$, in a form
%\begin{align}
%    i = i_{0}\left(\left\{a_{i}\right\}, \left\{c_{i}\right\}, \dots\right)f\left( \eta, \left\{a_{i}\right\}, \left\{c_{i}\right\}, \dots\right).
%    \label{}
%\end{align}
%where $i_0$ is the exchange current density, and $f$ is some function describing SOMETHING?\@
The net reduction current, $i$, can be related to the intercalation flux, $j_i$, for a given species by the reaction stoichiometry.
For example, for lithium intercalation, $j_i = i/e$.
Extension to multiple reactions simply involves describing each $j_i$ as a sum over the relevant reactions.

It is worth noting that the reaction models currently implemented and discussed below follow the trend in battery modeling to neglect the impact of diffuse charge within double layers on the reaction kinetics.
Accounting for this involves using a model of the double layer to adjust the surface concentrations from those outside the double layer to those at the distance of closest approach to the electrode surface (the Stern layer), which actually drive the reaction, as well as accounting for the local electric field driving electron transfer.
These changes constitute the Frumkin correction~\cite{frumkin1933,bazant2005} to the reaction rate model, and have been recently reviewed~\cite{biesheuvel2009,yan2017theory}.
Frumkin-corrected Butler-Volmer reaction models have been applied to various electrochemical techniques including steady constant current~\cite{itskovich1977electric,kornyshev1981conductivity,bazant2005,biesheuvel2009}, voltage steps~\cite{streeter2008numerical}, current steps~\cite{vansoestbergen2010diffuse}, and linear sweep voltammetry~\cite{yan2017theory}, as well as nano~\cite{he2006dynamic} and porous~\cite{biesheuvel2011diffuse,biesheuvel2012electrochemistry} electrodes, with clear indication of departure from models neglecting double layers, especially at low salt concentrations with thick double layers (``Gouy-Chapman limit''~\cite{bazant2005,biesheuvel2009}).
To be used consistently with the models developed here, Frumkin reaction kinetics would need to be extended to concentrated electrolyte solutions, including models of individual ionic activities within the double layers, although Frumkin effects are reduced for very thin double layers at high salt concentration (``Helmholtz limit''~\cite{bazant2005,biesheuvel2009}).
On the other hand, Frumkin effects that dominate in dilute solutions could be important for practical battery operation at high rates, where severe electrolyte depletion can occur and limit the achievable power density.

\paragraph{Butler-Volmer kinetics}
Butler-Volmer reaction kinetics are described by exponential dependence on the activation overpotential, and the net reduction current can be written as
\begin{align}
    i = i_0\left( \exp{\left( -\frac{\alpha e\eta_{\mathrm{eff}}}{k_\mathrm{B}T} \right)}
    -\exp{\left( \frac{\left( 1-\alpha \right)e\eta_{\mathrm{eff}}}{k_\mathrm{B}T} \right)}\right)
    \label{eq:bv}
\end{align}
where $\alpha$ is a symmetry coefficient and $i_0$ is the exchange current density, the rate of reaction in the forward and reverse directions when the reaction is in equilibrium.
Depending on the system considered, the exchange current density could be modeled as constant~\cite{cogswell2015} or a function of species concentrations or activities.
Introduced in ref.~\cite{doyle1996}, the effective overpotential, $\eta_\mathrm{eff}$, accounts for any film resistance, $R_{\mathrm{film}}$, by
\begin{align}
    \eta_{\mathrm{eff}} = \eta + iR_\mathrm{film}.
    \label{eq:Rfilm}
\end{align}
Bazant and co-workers proposed a form for $i_0$ based on reacting species' activities and a transition state activity coefficient, $\gamma_\ddagger$~\cite{bai2011,cogswell2012,ferguson2012}, derived by assuming thermally activated transitions in an excess chemical potential energy surface with an electric field across the reaction coordinate contributing to the transition state~\cite{bazant2013},
\begin{align}
    i_0 = \frac{k_0ne{\left( a_{O}a_\mathrm{e}^n \right)}^{1-\alpha}a_{R}^\alpha}{\gamma_\ddagger}
    \label{eq:ecd_act}
\end{align}
where $k_0$ is a rate constant, $a_O = \prod_{i}a_i^{s_{i,O}}$, $a_e$ is the activity of the electrons (taken to be unity here), and $a_R = \prod_{j}a_j^{s_{j,R}}$.
The transition state activity coefficient is a postulate about the structure and characteristics of the transition state.
For example, for lithium intercalation into LiFePO$_4$, Bai et al.\ originally proposed $\gamma_\ddagger = {\left( 1 - c/c_{\mathrm{\max}} \right)}^{-1}$ where $c_{\mathrm{\max}}$ is the maximum concentration of lithium within the solid, to indicate the transition state excludes one site.
Of note, although the above expression is defined in terms of the activities of individual ions within the electrolyte, these quantities are difficult to directly measure.
Because we have not implemented models to estimate ion activities, the software currently assumes $a_{\mathrm{Li}^+} = \wt{c}_\ell$ for both electrolyte models.

It is also common to use an exchange current density form based solely on species concentrations~\cite{newman2004,thomas2002},
\begin{align}
%    I_0 = k_0c^{1-\alpha}c_s^\alpha{\left( c_{s,\mathrm{\max}} - c_s \right)}^\alpha.
%    I_0 = k_0\wt{c}^{\,1-\alpha}\wt{c}_{s,i}^{\:\alpha}{\left( 1 - \wt{c}_{s,i} \right)}^\alpha,
    i_0 = k_0\wt{c}_{\ell}^{\,1-\alpha}\wt{c}_{i}^{\:\alpha}{\left( 1 - \xi_i\wt{c}_{i} \right)}^\alpha,
    \label{eq:ecd_conc}
\end{align}
where $\wt{c}_{\ell} = c_{\ell}/c_{\ell,\mathrm{ref}}$, $\wt{c}_{i} = c_{i}/c_{s,\mathrm{ref}}$, and $\xi_i = c_{s,\mathrm{ref}}/c_{i,\mathrm{\max}}$, and the ref subscripts indicate some suitable concentration scale. For the case of lithium insertion, we will choose $c_{s,\mathrm{ref}} = c_{i,\mathrm{\max}}$, so $\xi_i = 1$.

\paragraph{Marcus-Hush-Chidsey kinetics}
Marcus-Hush electron transfer kinetics describe an electron transfer event in terms of a reaction coordinate corresponding to the collective rearrangement of species involved in the transition state~\cite{marcus1956oxidationreduction,marcus1965electron,hush1958adiabatic,marcus1993,hush1999electron,kuznetsov_book}.
The electron transfer event can occur when it is energetically equivalent to occupy the donor or acceptor species, and the fluctuations which lead to this are related to dielectric rearrangement of molecules and charges around the donor-acceptor pair.
The Franck-Condon condition is satisfied because the electron transfer event is much faster than the rearrangements of the nearby molecules contributing to the energy of the reacting species.
Following the overview of Fedorov and Kornyshev~\cite{fedorov2014}, the reductive and oxidative currents at an electrode can be calculated as an integral over the energy levels of the electrons in the electron conducting phase (the electrode),
\begin{align}
    i_\mathrm{red} &= ek^0c_{O}\int_{-\infty}^{\infty}\rho(z)n_e(z)W_\mathrm{red}\ud z
    \\
    i_\mathrm{ox} &= ek^0c_{R}\int_{-\infty}^{\infty}\rho(z)\left( 1-n_e(z) \right)W_\mathrm{ox}\ud z
    \label{}
\end{align}
where $k^0$ is some rate prefactor, $c_O = \prod_{i}c_i^{s_{i,O}}$ and $c_R = \prod_{j}c_j^{s_{j,R}}$, $z$ represents the energy level of electrons in the electrode, $\rho(z)$ is the density of states, and $n_e(z)$ is the Fermi function.
$W_\mathrm{red/ox}$ are the transition probabilities of the elementary reduction/oxidation electron transfer processes.
Marcus theory considers the collective motion and reorganization of molecules near the electron transfer event, causing higher or lower energy contributions to the initial or final states of the reaction event via the electrostatic interactions between nearby polarization and the electron donor/acceptor.
By treating this collective motion as approximately parabolic around their lowest energy configurations, the following can be derived for the transition probabilities,
\begin{align}
    W_\mathrm{red} &= k_\mathrm{w}\exp\left( -\frac{w_O}{k_\mathrm{B}T} \right)
    \exp\left( -\frac{{\left( \lambda + e\eta_f \right)}^2}{4\lambda k_\mathrm{B}T} \right)
    \\
    W_\mathrm{ox} &= k_\mathrm{w}\exp\left( -\frac{w_R}{k_\mathrm{B}T} \right)
    \exp\left( -\frac{{\left( \lambda - e\eta_f \right)}^2}{4\lambda k_\mathrm{B}T} \right),
    \label{}
\end{align}
where $k_\mathrm{w}$ is a prefactor with explicit dependence on various factors, including the overlap integrals for the wave functions of the various reaction elements~\cite{kuznetsov_book}.
We assume reaction symmetries by the equality of the reductive and oxidative prefactors and treat it as a lumped constant here.
$w_{R/O}$ are energies describing the probabilities of the species arriving and orienting at the reaction site.
$\lambda$ is the reorganization energy, related to the force constants or the curvature of the reaction parabolas, which we take to be the same for forward and reverse reactions, i.e.\ symmetric Marcus theory.
It is defined as the energy required to perturb the system to the stable configuration of the product without allowing electron transfer.
The driving force is the reaction change in excess free energy, $\Delta G_\mathrm{rxn}^\mathrm{ex}$,
\begin{align}
    e\eta_f = \Delta G_\mathrm{rxn}^\mathrm{ex} = e\eta + k_\mathrm{B}T\ln\left( \frac{\wt{c}_O}{\wt{c}_R} \right)
    \label{}
\end{align}
where we have assumed single-electron transfer, as supported by Marcus theory for elementary reaction events~\cite{bazant2013}.
The tildes indicate non-dimensional concentrations $\wt{c}_O = \prod_{i}\wt{c}_i^{s_{i,O}}$ and $\wt{c}_R = \prod_{j}\wt{c}_j^{s_{j,R}}$, with each scaled to its reference concentration ($c_{\ell,\mathrm{ref}}$ for ions in the electrolyte and $c_\mathrm{s,\mathrm{ref}}$ for intercalated species in the active material).
Thus, $\eta_f$ is also the departure of the electrode potential from the formal potential.
Interestingly, for non-electrode reactions in solution, this predicts a maximum in reaction rate at driving forces given by $\pm\lambda$, and experimental validation of this so-called ``inverted'' region of decreasing reaction rate at increasing driving force  (i.e.\ negative differential reaction resistance~\cite{bazant2017thermodynamic}) paved the way for the Nobel Prize of Marcus~\cite{marcus1993}.

Assuming $w_{R/O}$ are independent of the electron energy level (not necessarily true~\cite{kuznetsov_book,fedorov2014}), neglecting variation in the density of states, modifying the driving force to account for the energy levels of electrons along the Fermi distribution, and lumping constants into $k_M$, we can arrive at the Marcus-Hush-Chidsey (MHC) electron transfer reaction model~\cite{chidsey1991},
\begin{align}
    i_\mathrm{red/ox} = k_{M}c_{O/R}
    \exp\left( -\frac{w_{O/R}}{k_\mathrm{B}T} \right)
    \int_{-\infty}^{\infty}\exp\left( -\frac{{\left( z - \lambda \mp e\eta_f \right)}^2}{4\lambda k_\mathrm{B}T} \right)\frac{\ud z}{1 + \exp\left( z/k_\mathrm{B}T \right)},
    \label{}
\end{align}
where $z$ is related to the energy level of the electronic states in the metal, and integration is over the Fermi distribution.
Reduction/oxidation correspond to the top/bottom signs.

Curiously, Marcus-style kinetics were not used to describe electron transfer reactions in batteries until very recently~\cite{bai2014}, possibly in part because of the complexity of the expressions involving improper integrals, making their computational evaluation cumbersome.
But recently, various approaches have been developed to facilitate the evaluation of the MHC expression, including an asymmetric variant with different values of $\lambda$ for the initial and final states, referred to as ``asymmetric Marcus Hush'' (AMH) kinetics~\cite{oldham2011,bieniasz2012,migliore2012,zeng2014simple,zeng2015simple}.
We will focus exclusively on the symmetric variant here.
As a result, MHC kinetics are now as easy to simulate as Butler-Volmer kinetics, so we include them in this simulation software to facilitate comparison in porous electrode modeling and data fitting.
For reviews of (asymmetric) MHC kinetics and applications to experimental data, see refs.~\cite{henstridge2012marcus,laborda2013asymmetric}.

We make one final assumption following ref.~\cite{smith2017mhc_draft} before arriving at the form we will use.
The $w_{R/O}$ functions are related to the probabilities of the system arriving at the state described by the parabolic minima in the theory.
As in the derivation of the Butler-Volmer expression by Bazant~\cite{bazant2013}, we postulate that this could capture effects of concentrated solutions beyond simple probabilistic occupation related to the concentration prefactors above.
In other words, they are related to the excess chemical potential of the state corresponding to the parabolic minima.
Assuming a symmetric approach to the reacting state from the forward and reverse directions,
\begin{align}
    w_O = w_R = \mu_{M,\ddagger}^\mathrm{ex}
    \label{}
\end{align}
or
\begin{align}
    \exp\left( -\frac{w_{R/O}}{k_\mathrm{B}T} \right) \propto \frac{1}{\gamma_{M,\ddagger}}.
    \label{}
\end{align}
Thus, we arrive at our expressions for MHC kinetics,
\begin{align}
    i = i_M\left( \wt{c}_{O}k_\mathrm{red} - \wt{c}_{R}k_{\mathrm{ox}} \right)
    \label{}
\end{align}
with
\begin{align}
    k_\mathrm{red/ox} = \int_{-\infty}^{\infty}\exp\left( -\frac{{\left( z - \lambda \mp e\eta_f \right)}^2}{4\lambda k_\mathrm{B}T} \right)\frac{\ud z}{1 + \exp\left( z/k_\mathrm{B}T \right)}
    \label{eq:MHC_integral}
\end{align}
and
\begin{align}
    i_M = \frac{k_M}{\gamma_{M,\ddagger}}.
    \label{}
\end{align}
Although the derivation of the Butler-Volmer equation~\cite{bazant2013} and the approach followed above both lead to some prefactor related to the excess chemical potential of the transition state, we suggest that the two terms are not capturing the same physical phenomena and thus may differ in their functional forms.
In this approach to the microscopic Marcus theory, there is some separation between the energetic contributions accounted for in the microscopic reorganization and the ``approach'' contributions lumped into $\gamma_{M,\ddagger}$.
The Butler-Volmer derivation does not make this distinction, suggesting that the associated contributions in the two theories need not necessarily be equal.
%where the integration over $x$ corresponds to integrating over the Fermi distribution of energy levels of electrons in the electrode, $\lambda$ is the reaction reorganization energy~\cite{henstridge2012marcus}, reduction/oxidation corresponds to the top/bottom sign, and $\eta_f$ is the departure of the electrode potential from the formal potential which can be expressed in terms of the reaction overpotential,
%\begin{align}
%    \eta_f = \eta + \frac{k_\mathrm{B}T}{e}\ln\left( \frac{\wt{c}_O}{\wt{c}_R} \right).
%    \label{}
%\end{align}
%The non-dimensional concentrations indicate $\wt{c}_O = \prod_{i}\wt{c}_i^{s_{i,O}}$ and $\wt{c}_R = \prod_{j}\wt{c}_j^{s_{j,R}}$ with the tildes indicating scaling to some reference concentration at which the standard potential is defined.
%$i_M$ is some current rate prefactor, typically taken as a constant.
%However, the transitioning reaction coordinate described by the theory occurs once the system has arrived in a state corresponding to the parabolic minima, which could involve extra statistical or energetic contributions, leading to state dependence in $i_M$, including concentration dependence.
Finally, following Zeng et al.~\cite{zeng2014simple}, we replace Eq.~\ref{eq:MHC_integral} with
\begin{align}
    k_\mathrm{red/ox} \approx \frac{\sqrt{\pi\wt{\lambda}}}{1 + \exp\left( \pm\wt{\eta}_f \right)}
    \erfc\left( \frac{\wt{\lambda} - \sqrt{1 + \sqrt{\wt{\lambda}} + \wt{\eta}_f^2}}{2\sqrt{\wt{\lambda}}} \right)
    \label{}
\end{align}
where tildes indicate scaling by the thermal energy or voltage, $k_\mathrm{B}T$ or $k_\mathrm{B}T/e$, $\erfc(z) = 1 - \erf(z)$ is the complementary error function, and reduction/oxidation corresponds to the top/bottom sign.

\subsubsection{Species Conservation}
\label{sec:model_am}
Conservation of the intercalant within the solid particles should be specialized to describe the particular physics of the material being studied.
Several options are described below.
We will simulate the individual particles by describing neutral species transport within them and their mass exchange with the electrolyte via the electrochemical reactions.
This assumes that electron mobility within the active materials is much larger than that of the inserted species~\cite{thomas2002}.

\paragraph{Homogeneous}
When transport within the solid particles is fast, it can be computationally beneficial to approximate the particles with an average concentration, $\ovl{c}_s$~\cite{ferguson2014}.
Then, given a reacting surface area, $A_p$, and volume, $V_p$, the dynamics of intercalant $i$ can be described simply by the average intercalation rate from the electrochemical reaction, $\ovl{j}_p$,
\begin{align}
    \frac{\partial\ovl{c}_{i}}{\partial t} = \frac{A_p}{V_p}\ovl{j}_{p,i}.
    \label{}
\end{align}

\paragraph{Allen-Cahn Reaction}
For Allen-Cahn reaction particles, the reaction occurs as a volumetric source term.
This arises when particles are assumed to be either homogeneous or depth averaged in the direction normal to the reacting surface, as Bazant and co-workers have employed in models of LiFePO$_4$~\cite{bai2011,cogswell2012,li2014}.
Then, neglecting transport, the local rate of change of the concentration at particle location $\mathbf{r}$ is
\begin{align}
    \frac{\partial c_{i}}{\partial t} = \frac{A_p}{V_p}j_{p,i}\left(\mathbf{r}\right).
    \label{}
\end{align}

\paragraph{Cahn-Hilliard Reaction}
In this model, transport of the intercalant within the particles is described by species conservation,
\begin{align}
    \frac{\partial c_{i}}{\partial t} = -\dvg\mathbf{F}_{i}.
    \label{}
\end{align}
The flux can be modeled using linear irreversible thermodynamics~\cite{groot1962}, which postulates that fluxes arise from gradients in the diffusional chemical potential,
\begin{align}
    \mathbf{F}_{i} = -\frac{D_{i}}{\wt{T}}c_{i}\bnab\wt{\mu}_{i},
    \label{eq:Ngradmu}
\end{align}
where we have again used the Einstein relation.
The diffusional chemical potential is scaled to the thermal energy at some reference temperature, $k_\mathrm{B}T_\mathrm{ref}$, and can be calculated from Eq.~\ref{eq:muVarDef}.
Following Bazant~\cite{bazant2013}, $D_i$ can be rewritten in terms of tracer diffusivity in the dilute limit, $D_{0,i}$, the activity coefficient of species $i$, $\gamma_i = a_i/\wt{c}_i$, and the activity coefficient of the diffusion transition state, $\gamma_{\ddagger,i}^d$,
\begin{align}
    D_i = D_{0,i}\frac{\gamma_i}{\gamma_{\ddagger,i}^d}.
    \label{}
\end{align}
Assuming simple diffusion on a lattice in which the diffusion transition state has similar enthalpic contributions as the diffusing species in lattice sites but in which the transition state excludes two adjacent sites, $\gamma_i/\gamma_{\ddagger,i}^d = \left( 1-c_i/c_{i,\mathrm{\max}} \right)$, giving
\begin{align}
    \mathbf{F}_i = -\frac{D_{0,i}}{\wt{T}}c_i\left( 1-\frac{c_i}{c_{i,\mathrm{\max}}} \right)\bnab\wt{\mu}_i,
    \label{eq:CHRflux_lattice}
\end{align}
although other effects could be accounted for in $\gamma_{\ddagger,i}^d$ including stresses in the transition state~\cite{aziz1991}.
At the particle surface, the flux is given by the electrochemical reaction, $\wh{\mathbf{n}}\cdot\mathbf{F}_{i} = -j_{p,i}$ where $\wh{\mathbf{n}}$ is a unit normal vector pointing from the active material to the electrolyte.
The natural boundary condition~\cite{cahn1977,bazant2013} imposes a constraint on the concentration at the surface as a function of the surface energy, $\gamma_S$, $\wh{\mathbf{n}}\cdot\frac{\partial g}{\partial\bnab \wt{c}_{i}} = \wh{\mathbf{n}}\cdot\kappa\bnab\wt{c}_{i} = \frac{\partial\gamma_S}{\partial \wt{c}_{i}}$.

\paragraph{Solid Solution}
If the free energy can be described as a function of only the concentration (disregarding the effect of gradients and other contributions), then Eq.~\ref{eq:Ngradmu} can be rewritten as
\begin{align}
    \mathbf{F}_{i} = -\frac{D_{i}}{\wt{T}}c_{i}\frac{\partial\wt{\mu}_{i}}{\partial c_{i}}\bnab c_{i}
    = -\frac{D_{\mathrm{chem},i}}{\wt{T}}\bnab c_{i}
    \label{}
\end{align}
where $D_{\mathrm{chem},i} = D_{i}c_{i}\frac{\partial\wt{\mu}_i}{\partial c_i}$.
Here, it is sufficient to prescribe the flux at the surface, as in the Cahn-Hilliard reaction model, $\wh{\mathbf{n}}\cdot\mathbf{F}_{i} = -j_{p,i}$.

\subsection{Coupling Equations}
The general approach we take to simulate the two coupled phases is to simulate, within the same physical and simulated space as the electrolyte, a representative sample of particles, which are duplicated to the appropriate filling fraction of solids within the electrode.
Instead of simulating discrete particles, an alternative approach could involve directly simulating the distribution of particles at a given state using a population balance model~\cite{kumar1996solutionpart1,marchisio2005solution,ramkrishna2000population}.
This may improve accuracy at fixed computational cost, especially for simple particle models like the homogeneous approximation.
However, for the more complicated particle models which explore only a tiny fraction of their state space, the current approach may be more efficient.

For example, to simulate an electrode with a very narrow particle size distribution, we simulate one particle at each electrode position.
For wide distributions, we use multiple particles sampled randomly from a given input distribution, and at each location, multiple simulated particles interact with the same electrolyte.
The representative simulated particles at each electrode position are scaled to occupy the specified volume fraction of active material per electrode volume.
We allow the simulated particle size distribution to vary as a function of position, as this allows us to sample a broader distribution within the full electrode, but loses accuracy in situations with significant transport losses in the bulk electrolyte or electron-conducting phases.
When setting up simulations, the primary consideration is to have a good representation of the full particle size distribution within the macroscopic length scales of interest, which may be the full electrode at low currents or small regions at high currents with strong electrolyte depletion or electronic losses.

The two phases are coupled via the electrochemical reactions.
The volumetric reaction rate of species $i$, $R_{V,i}$, is related to the electrochemical reactions and the flux of that species out of the solid particles.
This can either be done by integrating the reaction rate over particle surfaces or by applying the divergence theorem to relate that to their average rate of filling.
For particle $p$, at a given position in the electrode,
\begin{align}
    \frac{\partial\ovl{c}_{p,i}}{\partial t} = \frac{1}{V_p}\int_{A_p}^{}j_{p,i}\ud A.
    \label{}
\end{align}
Thus, the volumetric source term can be cast as a sum over all the particles at a given position,
\begin{align}
    R_{V,i} = -\left( 1-\eps \right)P_L\sum_{p}\frac{V_p}{V_u}\frac{\partial\ovl{c}_{p,i}}{\partial t}
    \label{}
\end{align}
where $P_L$ is the loading percent of active material in the solid phase, $V_u = \sum_{p}V_p$, and $p$ indicates properties of particle $p$, and the summation is over particles at the location where $R_{V,i}$ is evaluated.

\subsection{Macroscopic Equations}
The overall current density per electrode area, $i_\mathrm{cell}$, is defined as the integral of the net charge consumed by the reactions in the electrodes per unit cross-sectional area,
\begin{align}
    i_\mathrm{cell} = \sum_{i}\int_{L_\mathrm{a}}^{}z_{i}eR_{V,i}\ud L
    = -\sum_{i}\int_{L_\mathrm{c}}^{}z_{i}eR_{V,i}\ud L
    \label{eq:Icell}
\end{align}
where $L_{\mathrm{a}}$ and $L_{\mathrm{c}}$ are the lengths of the anode and cathode.
We also find it useful to define a current in terms of a C-rate, which is determined by the electrode with the smaller capacity.
The capacity of the cell per area, $Q_A$, is given for the case of an electrode with a single intercalating species with maximum concentration $c_{\mathrm{\max},k}$ in electrode $k$ by the capacity per area of the limiting electrode,
\begin{align}
    Q_A = \min_{k\in\left\{ a,c \right\}}\left\{ eL_{k}\left( 1-\eps_k \right)P_{L,k}c_{\mathrm{\max},k} \right\}.
    \label{}
\end{align}
With that, we can define a C-rate current,
\begin{align}
    i_{\mathrm{cell,C}} = \frac{i_\mathrm{cell}}{Q_A}\tau_\mathrm{hr}
    \label{}
\end{align}
where $\tau_\mathrm{hr}$ is a conversion factor to ensure units of $\mathrm{hr}^{-1}$.

The overall utilization (state of charge) of electrode $k$ for species $i$, $u_{k,i}$, is defined in terms of the average filling fraction of all the particles in the electrode,
\begin{align}
    u_{k,i} = \frac{1}{L_k}\int_{L_\mathrm{k}}^{}\sum_{p}\frac{V_p}{V_u}\xi_i\wt{\ovl{c}}_{p,i}\ud L.
    \label{}
\end{align}
where $k$ indicates either the anode or cathode and the summation is evaluated as a function of position, given the selection of particles in the simulated electrode at that position.
From above, $\xi_i = c_{s,\mathrm{ref}}/c_{i,\mathrm{\max}}$, and $\wt{\ovl{c}}_{p,i} = \ovl{c}_{p,i}/c_{s,\mathrm{ref}}$, so the product is the average filling fraction of particle $p$.

The overall cell voltage, $\Delta\phi_{\mathrm{cell}} = \phi_c - \phi_a$ is defined as the difference in the electric potential at the cathode and anode current collectors.
There is an arbitrary datum for the potential, and we set $\phi_c = 0$ in the simulations.
We account for series resistance by defining an applied voltage, $\Delta\phi_{\mathrm{appl}}$, by
\begin{align}
    \Delta\phi_{\mathrm{cell}} = \Delta\phi_\mathrm{appl} - i_{\mathrm{cell}}R_{\mathrm{ser}},
    \label{}
\end{align}
where $R_\mathrm{ser}$ is the area specific resistance of the cell.

\section{Non-dimensional Equations}
\label{sec:ndim_eqns}
All times in the simulation are scaled to a reference time scale $t_\mathrm{ref}$.
We choose some representative diffusive time scale within the cell, $t_\mathrm{ref} = L_{\ell,\mathrm{ref}}^2/D_{\ell,\mathrm{ref}}$ where $L_{\ell,\mathrm{ref}}$ is a characteristic length within the cell (we will use the cathode length) and $D_{\ell,\mathrm{ref}}$ is a suitably chosen scale for the electrolyte diffusion coefficient.
%We also restrict the following to a binary electrolyte.

\subsection{Electrode Scale Equations}
\label{sec:ndim_electrode}
\subsubsection{Electrolyte Model}
Defining a reference electrolyte concentration, $c_{\ell,\mathrm{ref}}$ (e.g. $c_{\ell,\mathrm{ref}}/N_\mathrm{A} = 1\ \mathrm{M}$, where $N_A$ is the Avogadro constant),
\begin{align}
    \frac{\partial\left( \eps \wt{c}_{\ell,i} \right)}{\partial \wt{t}} &=
    \frac{1}{\nu_i}\Big(-\wtdvg\wt{\mathbf{F}}_{\ell,i} + \wt{R}_{V,i}\Big)
    \\
%    0 &= -\wtdvg\wt{\mathbf{i}}_{\ell} + \left( z_{+}\wt{R}_{V,+} + z_{-}\wt{R}_{V,-} \right)
    0 &= -\wtdvg\wt{\mathbf{i}}_{\ell} + \sum_{i}z_{i}\wt{R}_{V,i}
    \\
    \wt{\mathbf{i}}_{\ell} &= \sum_{i}z_{i}\wt{\mathbf{F}}_{\ell,i}
    \label{}
\end{align}
with $\wt{\bnab} = L_\mathrm{ref}\bnab$ and tildes indicating non-dimensionalization by the following scales
\begin{align}
%    \wt{\bnab} &= L_\mathrm{ref}\bnab
%    \\
    F_{\ell,\mathrm{ref}} &= \frac{c_{\ell,\mathrm{ref}}L_{\ell,\mathrm{ref}}}{t_\mathrm{ref}}
    \\
    R_{V,\mathrm{ref}} &= \frac{c_{\ell,\mathrm{ref}}}{t_\mathrm{ref}}
    \\
    i_{\ell,\mathrm{ref}} &= eF_{\ell,\mathrm{ref}}.
    \label{}
\end{align}
The (semi-)dilute fluxes are non-dimensionalized by
\begin{align}
    \wt{\mathbf{F}}_{\ell,i} &= -\frac{\eps}{\tau}\left(\wt{D}_{\ell,\mathrm{chem},i}\wt{\bnab} \wt{c}_{\ell,i} + \frac{\wt{D}_{\ell,i}}{\wt{T}}z_{i}\wt{c}_{\ell,i}\wt{\bnab}\wt{\phi}_{\ell}\right)
    \label{}
\end{align}
with diffusive scale as that chosen to define $t_\mathrm{ref}$ above.
The binary concentrated solution theory can be non-dimensionalized with
\begin{align}
    \wt{\mathbf{F}}_{\ell,+} &= -\frac{\nu_{+}\eps}{\tau}\wt{D}_{\ell}\wt{\bnab}\wt{c}_{\ell} + \frac{t_{+}^{0}\wt{\mathbf{i}_{\ell}}}{z_{+}}
    \label{}
    \\
    \wt{\mathbf{F}}_{\ell,-} &=  -\frac{\nu_{-}\eps}{\tau}\wt{D}_{\ell}\wt{\bnab}\wt{c}_{\ell} + \frac{t_{-}^{0}\wt{\mathbf{i}_{\ell}}}{z_{-}}
    \\
%    \wt{\mathbf{i}} &= -\frac{\eps\wt{\kappa}}{\tau}\left(\wt{\bnab}\wt{\phi}^r
    \wt{\mathbf{i}}_{\ell} &= -\frac{\eps\wt{\sigma}_{\ell}}{\wt{T}\tau}\left(\wt{\bnab}\wt{\phi}_{\ell}^r
    + \nu\wt{T}
    \left( \frac{s_+}{n\nu_+} + \frac{t_+^0}{z_{+}\nu_{+}} - \frac{s_{0}\wt{c}_{\ell}}{n\wt{c}_{\ell,0}} \right)
    \left( 1 + \frac{\partial\ln \gamma_\pm}{\partial\ln \wt{c}_{\ell}} \right)\wt{\bnab}\ln \wt{c}_{\ell}\right)
    \label{}
\end{align}
with conductivity scale
\begin{align}
%    \kappa_\mathrm{ref} = \frac{k_\mathrm{B}T}{e^2c_\mathrm{ref}D_\mathrm{ref}}.
    \sigma_\mathrm{ref} = \frac{e^2c_{\ell,\mathrm{ref}}D_{\ell,\mathrm{ref}}}{k_\mathrm{B}T_\mathrm{ref}}.
    \label{}
\end{align}

\subsubsection{Solid phase electronic model}
The solid phase current density and conductivity are scaled to $i_\mathrm{ref}$ and $\sigma_\mathrm{ref}$ respectively such that
\begin{align}
    0 = -\wtdvg\wt{\mathbf{i}}_s - \left( z_{+}\wt{R}_{V,+} + z_{-}\wt{R}_{V,-} \right)
    \label{}
\end{align}
and
\begin{align}
    \wt{\mathbf{i}}_s = -\frac{\left( 1-\eps \right)}{\tau}\wt{\sigma}\wt{\bnab}\wt{\phi}_s.
    \label{}
\end{align}
%with
%\begin{align}
%    \sigma_\mathrm{ref} = \frac{i_{s,\mathrm{ref}}L_\mathrm{ref}e}{k_\mathrm{B}T}.
%    \label{}
%\end{align}

\subsection{Single Particle Equations}
\label{sec:ndim_single}
In the solid particles, we will use the same time scale as in the electrode scale equations but different length and concentration scales,
\begin{align}
    F_{s,\mathrm{ref}} &= \frac{L_{s,\mathrm{ref}}c_{s,\mathrm{ref}}}{t_\mathrm{ref}}
%    \\
%    L_{s,\mathrm{ref}} &= L_p
    \label{}
\end{align}
where $L_{s,\mathrm{ref}}$ and $c_{s,\mathrm{ref}}$ are relevant scales to the particle. We will use $L_{s,\mathrm{ref}} = L_p$, where $L_p$ is a characteristic length scale of the active material particle and which may vary by particle. For the case of lithium insertion electrodes with a single intercalating species with maximum concentration given by $c_\mathrm{\max}$, we choose the solid reference concentration to be the maximum filling for lithium insertion cells, $c_{s,\mathrm{ref}} = c_{\mathrm{\max}}$.

\subsubsection{Electrochemical Reactions}
The various forms of the overpotential are all scaled to the thermal voltage, $\frac{k_\mathrm{B}T_\mathrm{ref}}{e}$, and the rate prefactors and current density are both scaled to
\begin{align}
%    k_{0,\mathrm{ref}} &= \frac{eL_{s,\mathrm{ref}}c_{s,\mathrm{ref}}}{t_\mathrm{ref}}
    i_{s,\mathrm{ref}} &= eF_{s,\mathrm{ref}}.
    \label{}
\end{align}
The film resistance is scaled to $R_\mathrm{film,ref} = \frac{k_\mathrm{B}T_\mathrm{ref}}{ei_{s,\mathrm{ref}}}$, so
\begin{align}
    \wt{\eta}_\mathrm{eff} = \wt{\eta} + \wt{i}\wt{R}_\mathrm{film}
    \label{}
\end{align}
and the Butler-Volmer expression is
\begin{align}
    \wt{i} = \wt{i}_0\left( \exp\left( -\alpha\wt{\eta}_\mathrm{eff}/\wt{T} \right)
    - \exp\left( \left( 1-\alpha \right)\wt{\eta}_\mathrm{eff}/\wt{T} \right)\right)
    \label{}
\end{align}
while the MHC expression is
\begin{align}
    \wt{i} = \wt{i}_M\left( \wt{c}_{O}k_\mathrm{red} - \wt{c}_{R}k_\mathrm{ox} \right).
    \label{}
\end{align}

\subsubsection{Species Conservation}
We scale the rate of intercalation of each species in particle $p$, $j_{p,i}$, to the active material reference flux, $N_{s,\mathrm{ref}}$, such that, for homogeneous particles,
\begin{align}
    \frac{\partial\wt{\ovl{c}}_{i}}{\partial\wt{t}} = \delta_{p,L}\wt{\ovl{j}}_{p,i}
    \label{}
\end{align}
where $\wt{\ovl{c}}_i = \ovl{c}_i/c_{s,\mathrm{ref}}$, and
\begin{align}
    \delta_{p,L} = \frac{A_{p}L_{p}}{V_{p}}.
    \label{}
\end{align}
For Allen-Cahn reaction particles,
\begin{align}
    \frac{\partial\wt{c}_{i}}{\partial\wt{t}} = \delta_{p,L}\wt{j}_{p,i}\left(\, \wt{\mathbf{r}} \,\right)
    \label{}
\end{align}
where $\wt{\mathbf{r}}$ is non-dimensionalized by $L_p$.
For Cahn-Hilliard reaction particles,
\begin{align}
    \frac{\partial \wt{c}_{i}}{\partial \wt{t}} = -\wtdvg\wt{\mathbf{F}}_{i}.
    \label{}
\end{align}
The non-dimensional flux is given by
\begin{align}
    \wt{\mathbf{F}}_{s,i} = -\frac{\wt{D}_{i}}{\wt{T}}\wt{c}_{i}\wt{\bnab}\wt{\mu}_{i}
    \label{}
\end{align}
where the diffusivity is scaled to
\begin{align}
    D_{s,\mathrm{ref}} = \frac{L_{s,\mathrm{ref}}^2}{t_\mathrm{ref}}.
    \label{}
\end{align}
For the case of diffusion on a lattice with a simple excluded site model for the transition state, Eq.~\ref{eq:CHRflux_lattice} becomes
\begin{align}
    \wt{\mathbf{F}}_i = -\frac{\wt{D}_{0,i}}{\wt{T}}\wt{c}_i\left( 1-\xi_i\wt{c}_i \right)\wt{\bnab}\wt{\mu}_i.
    \label{}
\end{align}
The natural boundary condition is scaled using
\begin{align}
    \kappa_\mathrm{ref} &= k_\mathrm{B}T_{\mathrm{ref}}L_{s,\mathrm{ref}}^2c_{s,\mathrm{ref}}
    \\
    \gamma_{S,\mathrm{ref}} &= \frac{\kappa_{\mathrm{ref}}}{L_{s,\mathrm{ref}}}
    \label{}
\end{align}
such that
\begin{align}
    \wh{\mathbf{n}}\cdot\wt{\bnab}\wt{c}_{i} = \frac{1}{\wt{\kappa}}\frac{\partial\wt{\gamma}_S}{\partial\wt{c}_{s}}
    \label{}
\end{align}
at the particle surface, and the flux boundary condition is simply
\begin{align}
    \wh{\mathbf{n}}\cdot\wt{\mathbf{F}}_{i} = -\wt{j}_{p,i}.
    \label{}
\end{align}
For solid solution particles,
\begin{align}
    \wt{\mathbf{F}}_{i} = -\frac{\wt{D}_{s,\mathrm{chem},i}}{\wt{T}}\wt{\bnab}\wt{c}_{s,i}
    \label{}
\end{align}
with the same flux boundary condition as the Cahn-Hilliard reaction particles.

\subsection{Coupling Equations}
Here, we retain the same scales as above, such that
\begin{align}
    \frac{\partial\wt{\overline{c}}_{p,i}}{\partial\wt{t}}
    = \delta_{p,L}\int_{\wt{A}_p}^{}\wt{j}_{p,i}\ud\wt{A}
    \label{}
\end{align}
and
\begin{align}
    \wt{R}_{V,i} = -\beta\left( 1-\eps \right)P_L\sum_p\frac{V_p}{V_u}\frac{\partial\wt{\overline{c}}_{p,i}}{\partial\wt{t}}
    \label{}
\end{align}
with $\beta = c_{s,\mathrm{ref}}/c_{\ell,\mathrm{ref}}$.

\subsection{Macroscopic Equations}
Keeping the same scales as in the electrode model,
\begin{align}
    \wt{i}_\mathrm{cell} = \sum_i\int_{\wt{L}_a}^{}z_i\wt{R}_{V,i}\ud\wt{L}
    = -\sum_i\int_{\wt{L}_c}^{}z_i\wt{R}_{V,i}\ud\wt{L}
    \label{}
\end{align}
and
\begin{align}
    u_{k,i} = \frac{1}{\wt{L}_k}\int_{\wt{L}_\mathrm{k}}^{}\sum_{p}\wt{V}_p\xi_i\wt{\ovl{c}}_{p,i}\ud \wt{L}.
    \label{}
\end{align}
where $k$ indicates either the anode or cathode, $\wt{L}_k = L_k/L_{\ell,\mathrm{ref}}$, and $\wt{V}_p = V_p/V_u$.
The series resistance can be scaled by the electrode model scales, $R_\mathrm{ser,ref} = \frac{k_\mathrm{B}T_\mathrm{ref}}{ei_\mathrm{ref}}$, and the potentials can all be scaled by the thermal voltage at the reference temperature, $k_\mathrm{B}T_\mathrm{ref}/e$, so
\begin{align}
    \Delta\wt{\phi}_\mathrm{cell} = \Delta\wt{\phi}_\mathrm{appl} - \wt{i}_\mathrm{cell}\wt{R}_\mathrm{ser}.
    \label{}
\end{align}

\section{Model implementation}
\label{sec:implementation}
To solve the system of PDE's in the model, we take the general approach of discretizing each in space using some variant of the finite volume method to obtain a system of differential algebraic equations (DAE's), and then stepping in time using a variable-order adaptive time stepper.
We discretize in space using finite volume methods both for their robustness to steep gradients and also their mass conservation to within numerical accuracy~\cite{eymard2007,moukalled2016}.
We use the IDAS time stepper of the SUNDIALS integration suite~\cite{hindmarsh2005} to solve the resulting DAE's with a backward differentiation formula approach.
The model and simulation are defined within the DAE Tools software package~\cite{nikolic2016}, which provides a modeling language environment within Python similar to that of more specialized modeling languages like gProms~\cite{barton1994modeling} or Modelica~\cite{fritzson1998modelica}.
This allows use of the full depth of the general purpose Python environment while adding helpful concepts from modeling languages such as logical separation of model definition from simulation setup and equation oriented model definition.

For example, to define a model, the user writes the spatially discretized equations in a familiar form within a model class and defines simulation particulars such as parameter values and initial conditions elsewhere.
The DAE Tools software handles the formation of all underlying system matrices and interactions with other numerical libraries involved in the actual time advancement.
As an extra convenience, it wraps IDAS with the ADOL-C automatic differentiation library~\cite{griewank1996} to form the analytical-accuracy Jacobian matrix, greatly facilitating the solution of non-linear systems of equations involved in the implicit time stepping.
This eases additions and modifications to the model, which can be made without user input of analytical derivatives or reliance on numerical approximation of the Jacobian.
This approach enables MPET developers to work entirely within a Python environment while using the fast and vetted lower level numerical libraries for computationally expensive or analytically tedious aspects of the simulation.

\subsection{Software Organization and Structure}
\begin{figure}[h]
    \centering
    \includegraphics[width=0.4\textwidth]{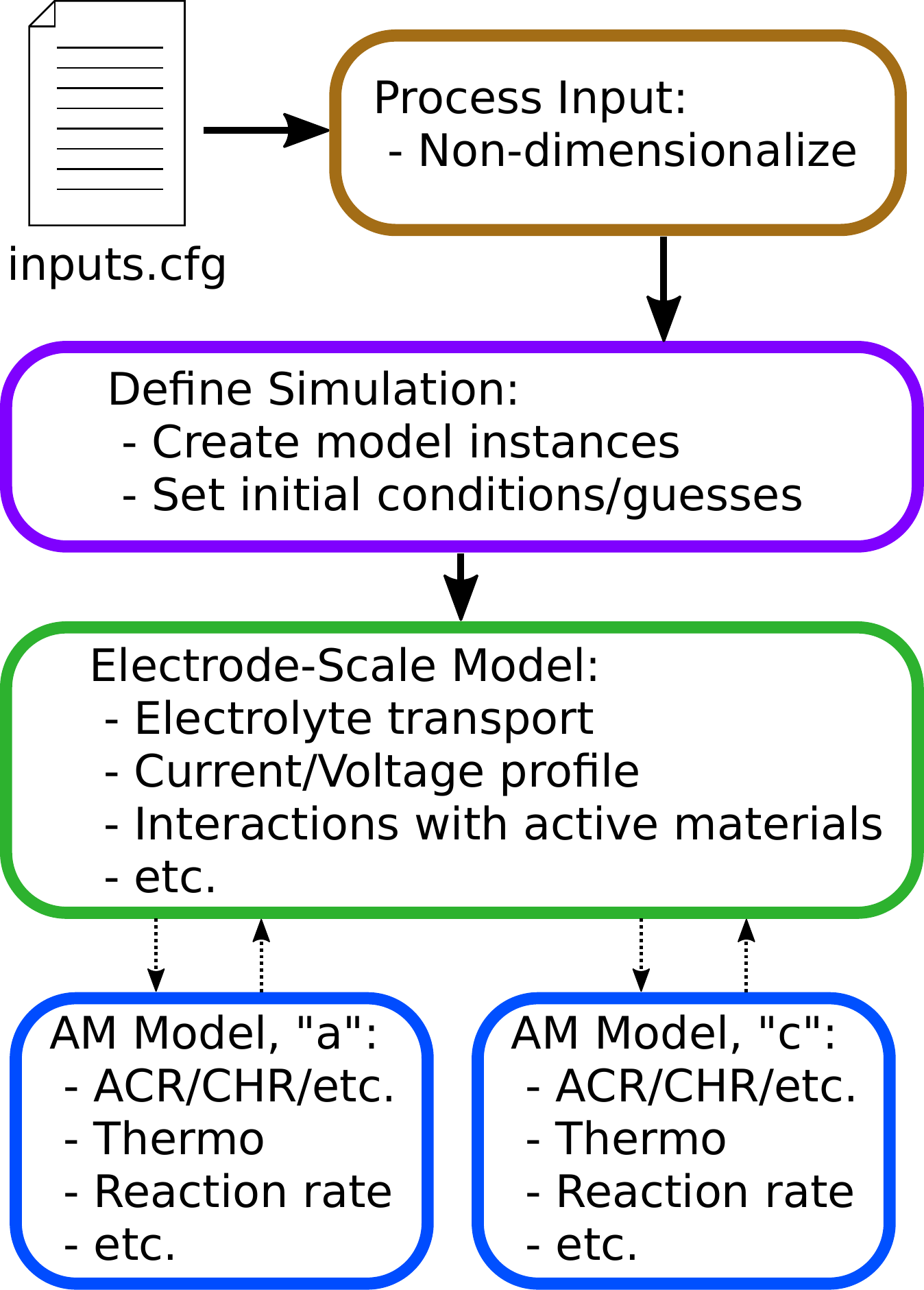}
    \caption{Structure of software. Simulation definition and input handling is separated from model equation definitions, and models are separated into electrode scale models and active material (AM) models which exchange required information.}
    \label{fig:software_structure}
\end{figure}

To make MPET flexible enough to simulate arbitrary active material models in the context of either porous electrodes or single particles to study their dynamics directly, we chose to develop the software to logically and structurally isolate the material models from the overall cell model using an object oriented framework, see Figure~\ref{fig:software_structure}.
Objects containing local information about active material models exchange only the required information with the cell model, reducing assumptions built into each section of the software and leading to a more modular, extensible structure.
Parameters are specified within input files using a standard config-file syntax for ease of scripting multiple simulations.
These input files are split according to information used by the overall system (e.g.\ specified current profile and system dimensions) and those defining properties of the simulated active materials (e.g.\ material thermodynamics, transport, and reaction properties).
This enables a user to reference standard material property files while editing only a system input file related to cell design.
These inputs are then non-dimensionalized for the simulation, as described in Section~\ref{sec:ndim_eqns}, and passed to the system model.

Depending on the simulation to be carried out (e.g.\ perfect electrolyte bath, half-cell, full-cell, with or without particle size variability), the system model creates as many instances of the appropriate active material model as necessary for the representative active material particles and establishes communication between them to only exchange key information.
The active material particles get access only to local concentrations and potentials in the bulk phases near the particular active material particle, and the bulk surrounding phases only need access to the total integrated reaction rate of each particle.
This isolation makes it relatively straightforward to extend MPET's capabilities by adding new material models or modifying existing ones to add relevant physics such as stresses and strains~\cite{garcia2004,christensen2006,vanderven2009,cogswell2012,bazant2013,sethuraman2012realtime,ulvestad2014single} or hierarchical structures~\cite{dargaville2010predicting}.
For example, to add and simulate a new model, a user must define the non-dimensional equations as a model class, specify the initial conditions via the simulation class, and update the handling of parameters to read and non-dimensionalize any new inputs.
Material models can easily use distinct discretization and/or geometries such as the homogeneous or CHR particles.
We have also already included the two-repeating-layers model developed and applied to single graphite particles~\cite{guo2016,smith2017_intercalationdraft}.
Other variations such as 2D particles of arbitrary geometry discretized using the finite volume~\cite{dargaville2015leastsquares} or finite element method would also be straightforward, especially with the DAE Tools interface to the deal.ii finite element library~\cite{nikolic2016,bangerth2007dealii}.
The structure also makes it a straightforward addition to implement multiple particle types to be simulated within each electrode region.
Finally, it makes it easier to modify specific parts of any given model with confidence that side-effects will be minimized through the isolation of the logical parts.

In order to facilitate good reproducibility of scientific computations, MPET defaults to storing each simulation output within time-stamped directories, along with both input files and a snapshot of the source code which ran that simulation.
Users can disable this feature to use their own system or take advantage of it by keeping a log matching the time-stamped directories with notes about each simulation.
Simulation outputs are stored by default in a binary format which is readable using common scientific computing software including Python (with SciPy), R, and MATLAB\@.
MPET includes a script to perform some basic plotting using this output and also a script to convert the output to comma-separated value (CSV) text files, which can be nearly universally interpreted.

\subsection{General Options}
As discussed above, the software's structure makes it flexible enough for a range of possible simulation options, and the key options are described here.
Either the current or voltage can be imposed as piecewise-constant functions with arbitrary numbers of steps (simulated with fast but continuous steps, similar to the method proposed in ref.~\cite{boovaragavan2010continuum}).
With minor changes to the code, completely arbitrary functions are possible.
Currents are imposed relative to a $1\ \mathrm{C}$ (dis)charge based on the simulated battery's limiting electrode, and optional cut-off voltages terminate the simulation if the system voltage reaches their values.
Because it is sometimes convenient to continue an old simulation, this can be done by specifying the location of a stored output.
Discretization is specified in terms of the number of finite volumes in each region of the electrode.
If the anode is set to have zero volumes, the simulation uses a flat lithium metal counter-electrode with specified Butler-Volmer reaction rate kinetics.
If only a single cathode volume is simulated, it places any simulated particles within a perfect electrolyte bath without any transport limitations for either single- or multi-particle studies neglecting electrolyte and bulk electrode losses.
Within each discretized porous electrode volume, the chosen number of particles are simulated, drawing from a specified log-normal size distribution.
Conductivity losses in the bulk electron-conducting phase or between individual particles within each volume of simulated electrodes are optionally neglected or simulated as described in Section~\ref{sec:model_bulk_e_losses}.
Electrode lengths, electrolyte volume fractions, loading percents (volume fraction of active material within the electron-conducting phase), and Bruggeman exponents describe the geometry characteristics of effective transport within the porous electrolyte.
The electrolyte can be specified either as a dilute model or using a full Stefan-Maxwell concentrated solution theory model as described in Section~\ref{sec:model_elyte}.
Arbitrary functions can be specified for the electrolyte transport properties by defining them as Python functions.

Active material particles are specified by the basic model for their transport processes (e.g. Fickian diffusion, homogeneous, Cahn-Hilliard reaction), the thermodynamic function describing them (also specified as arbitrary Python functions), and the remaining properties associated with the particular model.
The specified transport model determines the type of material model (defining the discretization method and equations solved) used in the simulations.
When applicable, the transport flux prefactor can be specified as an arbitrary function.
Electrochemical reactions are defined as Butler-Volmer, Marcus-Hush-Chidsey, or (experimental) Marcus kinetics as formulated by Bazant~\cite{bazant2013}, and a reaction film resistance can be added to any of these.
The Butler-Volmer exchange current density is taken as one of a few built-in options, as outlined in Section~\ref{sec:echemRxns}, but can be specified as arbitrary functions within the particle model source code.
To facilitate stability of some models, we provide the options to replace the most extreme regions of the log terms of ideal-solution parts of thermodynamic models with linear extrapolations.
Finally, to ensure that materials phase separate when they naturally would based on minor fluctuations, we include the option to add Langevin noise to the rate of change of concentration within each solid finite volume~\cite{cogswell2012}.

\subsection{Numerical Methods}
For each of the methods detailed below, we will describe the discretization in non-dimensional form using the scales presented in Section~\ref{sec:ndim_eqns}. We will also drop subscripts indicating species and location (electrolyte or active material particle) for clarity with discretization subscripts.
\subsubsection{Electrolyte}
The electrolyte equations are discretized using a typical 1D finite volume scheme with cell centers~\cite{moukalled2016}.
They are non-dimensionalized using the same scales as in the electrode scale equations, Section~\ref{sec:ndim_electrode}.
For example, for species conservation in cell $j$ with width $\Delta\wt{x}$ and neglecting subscripts for electrolyte/species identifiers for clarity,
\begin{align}
    \frac{\partial\wt{c}_j}{\partial\wt{t}} = \frac{\wt{F}_{j-1/2} - \wt{F}_{j+1/2}}{\Delta\wt{x}}
    \label{}
\end{align}
where the $j$ subscript represents an average quantity within the discretized volume, and the $F_{j\pm 1/2}$ represent components of the flux normal to the faces on the right/left of cell $j$ in the positive $x$ direction, where $x$ is along the length of the cell going between the current collectors.
The flux is approximated using a finite difference two-point formula based on the adjacent cell centers.
%For uniform discretizaiton, this leads to second order convergence in mesh spacing.
%At the interfaces between the separator and electrodes, where the discretization is not generally equal on both sides of the interface, this leads to a first order approximation.
Where required (e.g.\ for electromigration terms), face values of field variables such as concentration are approximated using harmonic means, which greatly enhances stability in regions of strong electrolyte depletion.
The harmonic mean also better represents variation in transport prefactors~\cite{russell1983finite}.

\subsubsection{Active Material}
The active material can be defined as spherical, cylindrical, or a rectangular grid approximation of platelet-like particles common in LiFePO$_4$~\cite{smith2012columnar}.
Those on a rectangular grid are discretized like the electrolyte, and the ratio of reacting area to volume is calculated as a function of the length and thickness, $h_p$, of the particle assuming reactions only occur on the top and bottom surfaces~\cite{li2014}.
For spheres and cylinders, the ratio of particle volume to reacting area is specified fully by the particle radius.
For cylinders, we use the particle thickness, $h_p$, for clarity in this section.
The spherical and cylindrical particles are discretized using a variant of finite volumes directly taken from Zeng et al.~\cite{zeng2013,zeng2014}.
They are non-dimensionalized using the same scales as in the single particle scale equations, Section~\ref{sec:ndim_single}.
For the cylindrical particles, we follow the same method as that in Zeng and Bazant, but modify it for the cylindrical geometry by changing the calculation of the volumes and areas of each sub-shell.
That is, for both geometries, the systems of equations can be represented as
\begin{align}
    \wt{\mathbf{M}}\wt{\mathbf{V}}\frac{\partial\wt{\mathbf{c}}}{\partial \wt{t}} = \wt{\mathbf{b}},
    \label{}
\end{align}
where $\wt{\mathbf{c}}$ is a vector of concentrations at positions going from the center of the particle to the surface of the particle with $N$ sub-volumes, $\wt{\mathbf{M}}$ is a tridiagonal matrix given by
\begin{align}
    \wt{\mathbf{M}} &= \begin{bmatrix}
        \frac{3}{4} & \frac{1}{8} & 0 & 0 & \cdots & 0 & 0 & 0 & 0 \\
        \frac{1}{4} & \frac{3}{4} & \frac{1}{8} & 0 & \cdots & 0 & 0 & 0 & 0 \\
        0 & \frac{1}{8} & \frac{3}{4} & \frac{1}{8} & \cdots & 0 & 0 & 0 & 0 \\
        \vdots & \vdots & \vdots & \vdots & \ddots & \vdots & \vdots & \vdots & \vdots \\
        0 & 0 & 0 & 0 & \cdots & \frac{1}{8} & \frac{3}{4} & \frac{1}{8} & 0 \\
        0 & 0 & 0 & 0 & \cdots & 0 & \frac{1}{8} & \frac{3}{4} & \frac{1}{4} \\
        0 & 0 & 0 & 0 & \cdots & 0 & 0 & \frac{1}{8} & \frac{3}{4} \\
    \end{bmatrix},
    \label{}
\end{align}
and $\wt{\mathbf{V}}$ is a diagonal matrix defined by the volumes of the shells scaled to $L_p^3$.
Indexing the volumes with $j=1\dots N$, and with a uniformly spaced radial vector, $\wt{r}_j$, with $N$ points going from the particle center to its surface, then for a sphere $\wt{\mathbf{V}}$ has components defined by
\begin{align}
    \wt{V}_{jj}^\mathrm{sphere} = \begin{cases}
        4\pi \left( \frac{\Delta\wt{r}^3}{24} \right), & j = 1 \\
        4\pi \left( \wt{r}_j^2\Delta\wt{r} + \frac{\Delta \wt{r}^3}{12} \right), & j = 2\dots N-1 \\
        4\pi\left( \frac{\wt{r}_j^3}{3} - \frac{{\left( \wt{r}_j - \Delta\wt{r}/2 \right)}^3}{3} \right), & j = N \\
    \end{cases},
    \label{}
\end{align}
and for a cylinder,
\begin{align}
    \wt{V}_{jj}^\mathrm{cylinder} = \begin{cases}
        \pi\wt{h}_p \frac{\Delta\wt{r}^2}{4}, & j = 1 \\
        2\pi\wt{h}_p\wt{r}_j\Delta\wt{r}, & j = 2\dots N-1 \\
        \pi\wt{h}_p\left( \wt{r}_j\Delta\wt{r} - \frac{\Delta\wt{r}^2}{4} \right), & j = N \\
    \end{cases}.
    \label{}
\end{align}
The right hand side, $\wt{\mathbf{b}}$, is defined in relation to the radial components of the fluxes evaluated at the interfaces between the shells, $\wt{F}_{j\pm 1/2}$, and the areas of the interfaces between the shells, $\wt{A}_{j\pm 1/2}$.
The fluxes and areas are scaled to $F_{s,\mathrm{ref}}$ and $L_p^2$ respectively.
For each geometry,
\begin{align}
    \wt{b}_j = \wt{A}_{j-1/2}\wt{F}_{j-1/2} - \wt{A}_{j+1/2}\wt{F}_{j+1/2}.
    \label{}
\end{align}
The shell areas differ for the two geometries, given scaled to $L_p^2$,
\begin{align}
    \wt{A}_{1 - 1/2}^\mathrm{cylinder} &= \wt{A}_{1 - 1/2}^\mathrm{sphere} = 0,
    \\
    \wt{A}_{j + 1/2}^\mathrm{sphere} &= \begin{cases}
        4\pi{\left( \wt{r}_j + \frac{\Delta\wt{r}}{2} \right)}^2, & j = 1\dots N-1 \\
        4\pi \wt{r}_j^2, & j = N \\
    \end{cases},
    \\
    \wt{A}_{j + 1/2}^\mathrm{cylinder} &= \begin{cases}
        2\pi\wt{h}_p\left( \wt{r}_j + \frac{\Delta\wt{r}}{2} \right), & j = 1\dots N-1 \\
        2\pi\wt{h}_p \wt{r}_j, & j = N \\
    \end{cases}.
    \label{}
\end{align}
%The radius of the particle scaled to $L_p$ is denoted as $\wt{R}_p$.
The flux is calculated using the two-point centered finite difference approximation using the concentrations in the adjacent cells.
Unlike in Zeng et al.~\cite{zeng2013,zeng2014}, we use harmonic means for face value approximations of field variables as in the electrolyte.
This discretization scheme for the cylinders and spheres has the advantage of increased accuracy of simulated surface concentrations~\cite{zeng2013} while retaining mass conservation to within numerical precision.
It is slightly less robust than a more typical cell-centered finite volume scheme under certain situations, as the coupling between the adjacent volumes via the mass matrix can cause minor oscillations during the formation of very steep concentration gradients, but this typically only presents problems when beginning with very high currents from nearly full or empty particles.
These oscillations could be eliminated by using a flux limiter, commonly used in higher order discretization schemes and especially for hyperbolic problems~\cite{leveque2002finite}.
This approach also has the advantage, much like typical finite volume schemes, of naturally facilitating the application of flux boundary conditions, as applied to the active material at the particle center (from symmetry) and surface (from reaction).

When applied to CHR particles, to evaluate the diffusional chemical potential at grid points we must calculate the Laplacian of the concentration, and we follow Zeng and Bazant~\cite{zeng2014}.
For interior points, $j = 2\dots N-1$,
\begin{align}
    \wt{\nabla}^2\wt{c}_j^\mathrm{\,sphere}
    &= \frac{2}{\wt{r}_j}\left.\frac{\partial\wt{c}}{\partial\wt{r}}\right|_{\wt{r}_j}
    + \left.\frac{\partial^2\wt{c}}{\partial\wt{r}^2}\right|_{\wt{r}_j}
    \approx \frac{2}{\wt{r}_j}\frac{\wt{c}_{j+1} - \wt{c}_{j-1}}{2\Delta\wt{r}}
    + \frac{\wt{c}_{j-1} - 2\wt{c}_j + \wt{c}_{j+1}}{\Delta\wt{r}^2},
    \\
    \wt{\nabla}^2\wt{c}_j^\mathrm{\,cylinder}
    &= \frac{1}{\wt{r}_j}\left.\frac{\partial\wt{c}}{\partial\wt{r}}\right|_{\wt{r}_j}
    + \left.\frac{\partial^2\wt{c}}{\partial\wt{r}^2}\right|_{\wt{r}_j}
    \approx \frac{1}{\wt{r}_j}\frac{\wt{c}_{j+1} - \wt{c}_{j-1}}{2\Delta\wt{r}}
    + \frac{\wt{c}_{j-1} - 2\wt{c}_j + \wt{c}_{j+1}}{\Delta\wt{r}^2}.
    \label{}
\end{align}
For the center of the particle at $j = 1$, isotropy gives
\begin{align}
    \wt{\nabla}^2\wt{c}_1^\mathrm{\,sphere}
    &= 3\left.\frac{\partial^2\wt{c}}{\partial\wt{r}}\right|_{\wt{r}_1}
    \approx 3\frac{2\wt{c}_2 - 2\wt{c}_1}{\Delta\wt{r}^2},
    \\
    \wt{\nabla}^2\wt{c}_1^\mathrm{\,cylinder}
    &= 2\left.\frac{\partial^2\wt{c}}{\partial\wt{r}}\right|_{\wt{r}_1}
    \approx 2\frac{2\wt{c}_2 - 2\wt{c}_1}{\Delta\wt{r}^2}.
\end{align}
At the surface, we use a ghost point at $j = N+1$ and impose the concentration gradient within the Laplacian of the concentration,
\begin{align}
    \wh{\mathbf{n}}\cdot\wt{\bnab}\wt{c}_N
    &= \frac{1}{\wt{\kappa}}\frac{\partial\wt{\gamma}_S}{\partial\wt{c}_{s}}
    \approx \frac{\wt{c}_{N+1} - \wt{c}_{N-1}}{2\Delta \wt{r}}
    \label{}
\end{align}
giving $\wt{c}_{N+1} \approx \frac{2\Delta\wt{r}}{\wt{\kappa}}\frac{\partial\wt{\gamma}_S}{\partial\wt{c}_{s}} + \wt{c}_{N-1}$, such that
\begin{align}
    \wt{\nabla}^2\wt{c}_N^\mathrm{\,sphere}
    &\approx \frac{2}{\wt{r}_N}\frac{1}{\wt{\kappa}}\frac{\partial\wt{\gamma}_S}{\partial\wt{c}_{s}}
    + \frac{2\wt{c}_{N-1} - 2\wt{c}_N + \frac{2\Delta\wt{r}}{\wt{\kappa}}\frac{\partial\wt{\gamma}_S}{\partial\wt{c}_{s}}}{\Delta\wt{r}^2},
    \\
    \wt{\nabla}^2\wt{c}_N^\mathrm{\,cylinder}
    &\approx \frac{1}{\wt{r}_N}\frac{1}{\wt{\kappa}}\frac{\partial\wt{\gamma}_S}{\partial\wt{c}_{s}}
    + \frac{2\wt{c}_{N-1} - 2\wt{c}_N + \frac{2\Delta\wt{r}}{\wt{\kappa}}\frac{\partial\wt{\gamma}_S}{\partial\wt{c}_{s}}}{\Delta\wt{r}^2}.
    \label{}
\end{align}

\subsection{DAE Consistent Initial Conditions}
Because the discretized equations are coupled DAE's, we must begin the system from consistent initial conditions~\cite{brown1998consistent,ramadesigan2012}.
Although various approaches for initializing this type of model have been developed~\cite{wu2001initialization,boovaragavan2007quick,methekar2011perturbation}, we found it to be relatively robust to begin from a known equilibrium state and quickly ramp the applied current or voltage to the desired values.
Although this contributes to the stiffness of the system, the higher order, adaptive backward difference formula (BDF) time stepper we use mitigates the cost of this, obviating the need to use more sophisticated methods.

To run simulations with specified current, we begin at a zero-current state, where we can easily calculate the equilibrium potentials in each of the phases (because we also begin with uniform concentration profiles).
Similarly, for simulations with specified applied voltages, we begin from the calculated equilibrium applied voltage before ramping to the desired voltage.
In addition, to make the initialization more robust, we offset the diffusional chemical potentials in the solid phases throughout the simulation by their initial values, such that the initial equilibrium potentials everywhere throughout the simulation are zero.
We found this to substantially facilitate initialization compared to leaving diffusional chemical potential at their physical values, which can leave differences in potentials across the system on the order of a hundred thermal volts (e.g.\ a $3\ \mathrm{V}$ material against a lithium metal or graphite anode is at $\sim 100$ thermal volts compared to $k_\mathrm{B}T/e$ at room temperature).
Because these potentials appear within exponentials of rate expressions, offsetting them to begin the simulation at zero substantially facilitates the numerical determination of consistent initial conditions and has no impact on the simulation output upon post-applying the reverse-offset.
Finally, when performing a simulation continuation, we take the final state of the previous output as initial guesses for the new initial state.

\section{Examples}
\label{sec:examples}
As a detailed examination of all the applications and model variations of the software is beyond the scope of the present work, we instead focus here on a few examples highlighting some of the key distinguishing capabilities of MPET\@.

\subsection{Solid solutions and phase separation in porous electrodes}
\label{sec:ssps}
We begin with a comparison of solid solution and phase separating models within porous electrode simulations, which highlights the ability of the software to compare two different approaches to modeling the same material, variations of both of which are commonly used~\cite{baker2012,levi2005comparison}.
A complete comparison of the different models is beyond the scope of this work, so we present only a simplified example here as a demonstration of the model behavior.
To keep the system as simple as possible, we choose to investigate a fictitious material in which lithium intercalates and has a diffusional chemical potential described by a regular solution,
\begin{align}
    \mu_\mathrm{RS} = k_\mathrm{B}T\ln\left( \frac{\wt{c}}{1-\wt{c}} \right) + \Omega\left( 1-2\wt{c} \right) - \frac{\kappa}{c_\mathrm{\max}}\nabla^2\, \wt{c} + \mu^\Theta
    \label{eq:mu_testRS_ps}
\end{align}
where $\wt{c} = c/c_\mathrm{\max}$ is a filling fraction, $c_\mathrm{\max}$ is the maximum concentration of lithium in the active material, and with $\mu^\Theta$ arbitrarily set to $-2\ \mathrm{eV}$ defined relative to Li/Li$^+$.
The regular solution parameter, $\Omega$, is set to $3k_\mathrm{B}T_\mathrm{ref}$, where $T_\mathrm{ref} = 298\ \mathrm{K}$ is the absolute temperature at which the simulation is carried out.
We examine an electrode with particles with radius, $R = 1\ \mu\mathrm{m}$, and we choose the interfacial gradient penalty, $\kappa = 1.16\times10^{-7}\ \mathrm{J}/\mathrm{m}$ such that the phase interface width is several times the finite volume grid size, $20\ \mathrm{nm}$.
The maximum filling fraction in the active material, $c_\mathrm{\max} = 25\ \mathrm{M}$ is chosen arbitrarily.

\begin{figure}[h]
    \centering
    \includegraphics[width=0.4\textwidth]{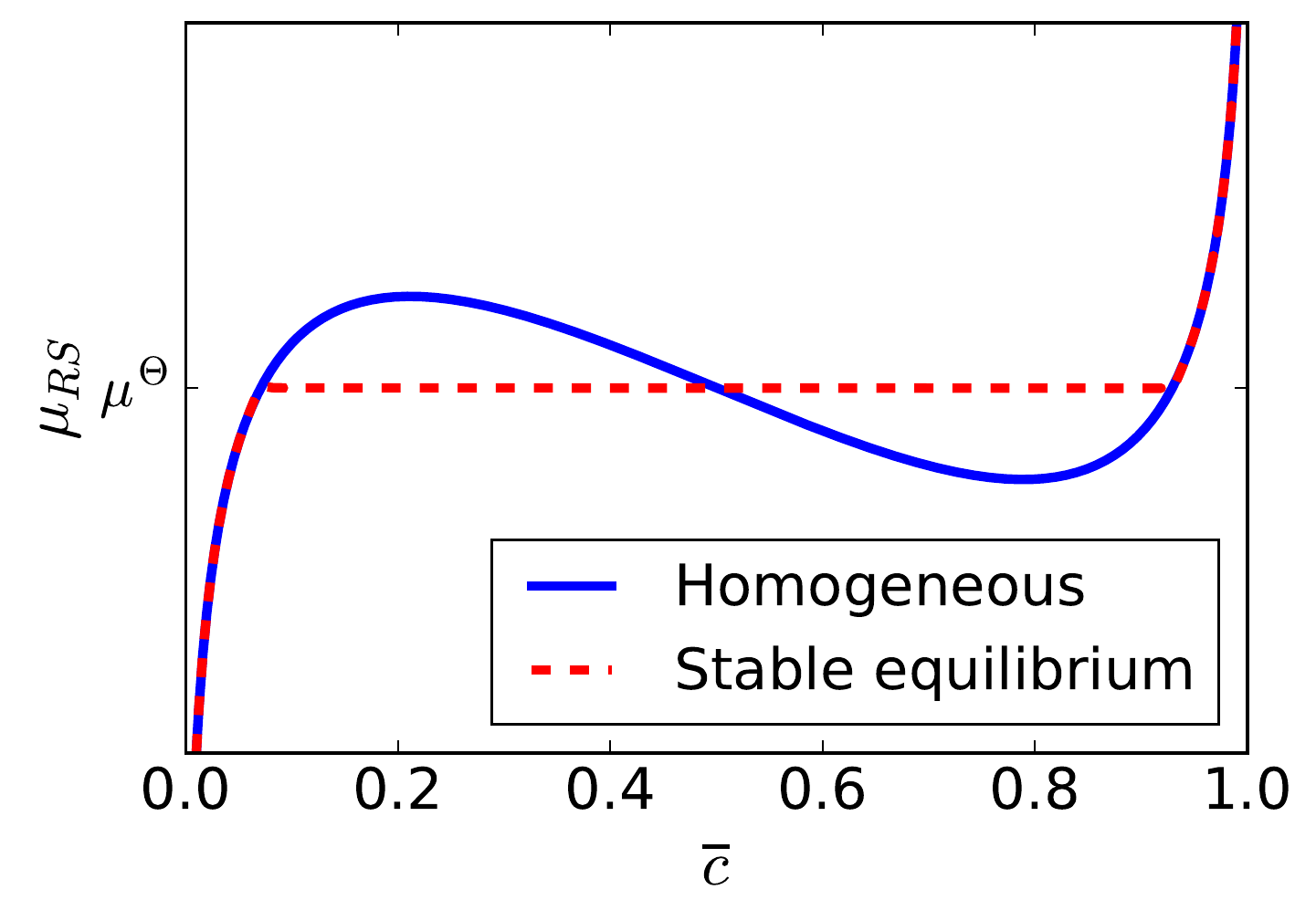}
    \caption{Diffusional chemical potential of a particle described by a regular solution as a function of average filling fraction, $\overline{c}$.}
    \label{fig:mu_ssps}
\end{figure}

To describe the thermodynamics of the solid solution material, we impose a diffusional chemical potential associated with the \emph{stable} equilibrium diffusional chemical potential defined by Eq.~\ref{eq:mu_testRS_ps}, a piece-wise continuous function defined as equal to $\mu_\mathrm{RS}$ outside the miscibility gap and $\mu^\Theta$ within it, the red dashed curve in Figure~\ref{fig:mu_ssps}.
Note, this could differ from a particular equilibrium filling/emptying path because a particle described by Eq.~\ref{eq:mu_testRS_ps} can enter metastable regions, leading to equilibrium hysteresis~\cite{levi1997simultaneous,dreyer2010,ferguson2014} which a solid solution model cannot capture without direction-dependent thermodynamic models.

\begin{figure}[h]
    \centering
    (a)
    \includegraphics[width=0.4\textwidth]{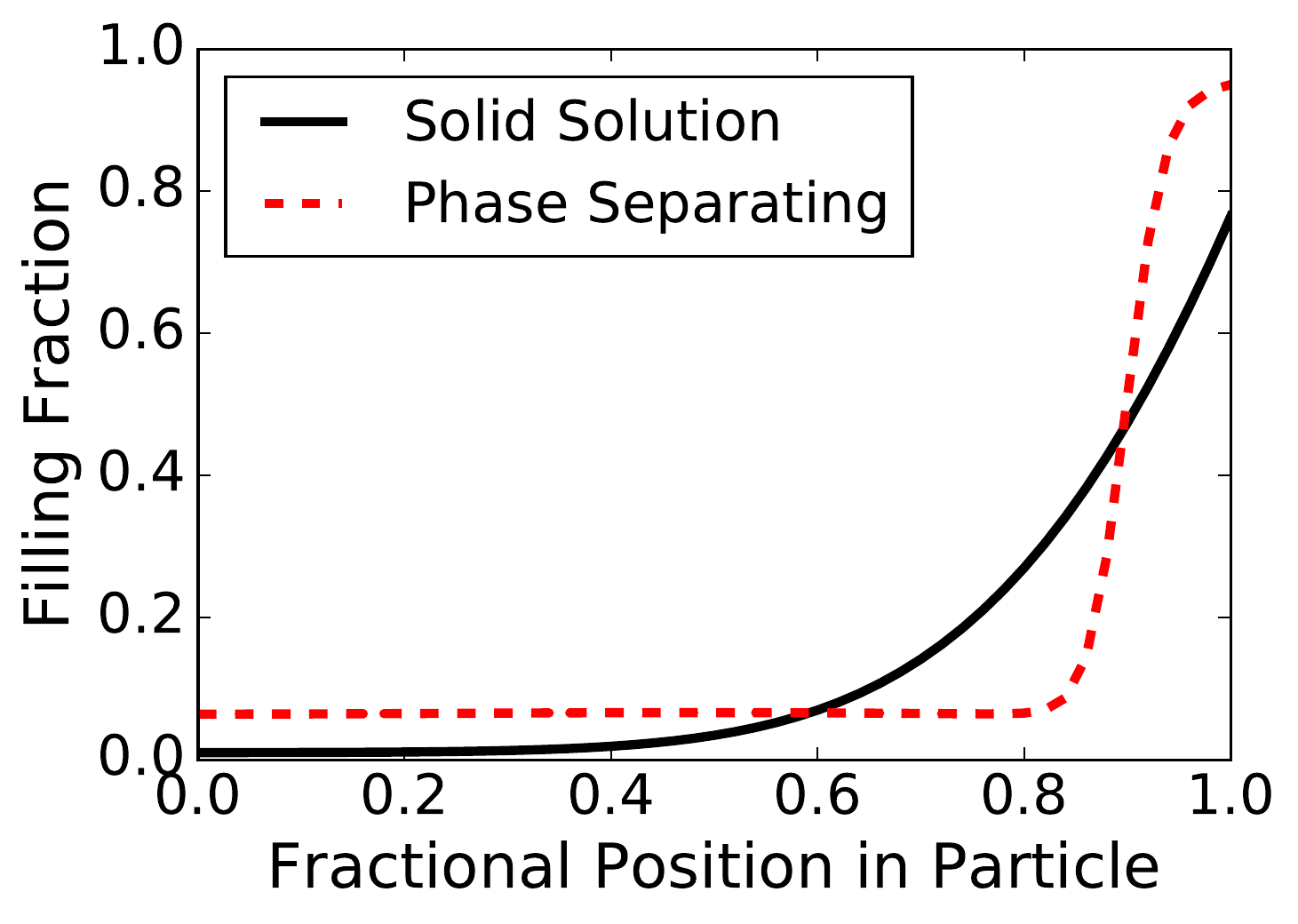}
    (b)
    \includegraphics[width=0.4\textwidth]{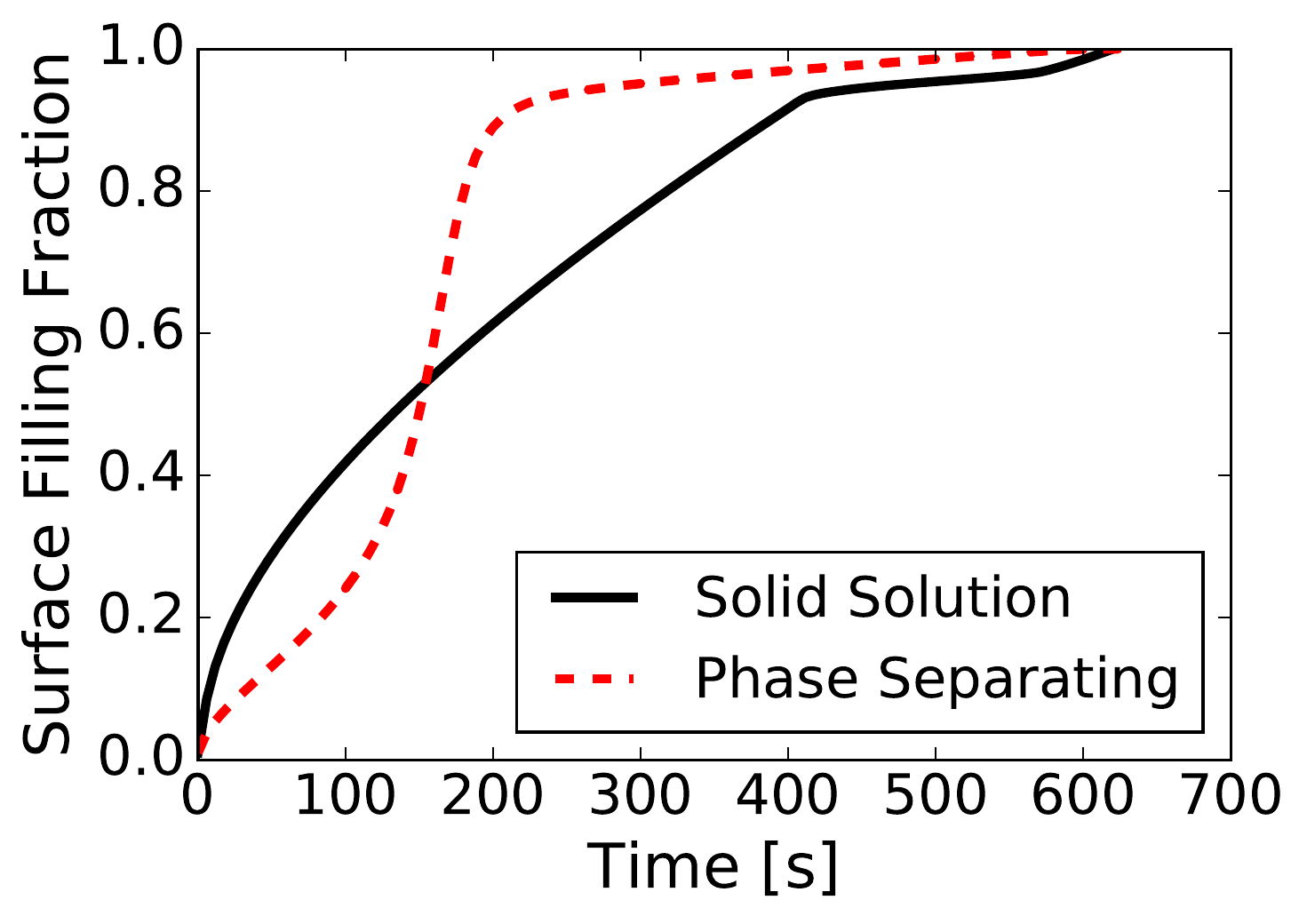}
    \caption{Concentration of particles nearest the separator in a $3\ \mathrm{C}$ discharge for both solid solution and phase separating models. In (a) we show a snapshot of the full concentration profile of the particle nearest the separator at $294\ \mathrm{s}$, and in (b) we show the surface concentration of that particle as a function of time.}
    \label{fig:ssps_csld_3C}
\end{figure}

\begin{figure}[h]
    \centering
    (a)
    \includegraphics[width=0.4\textwidth]{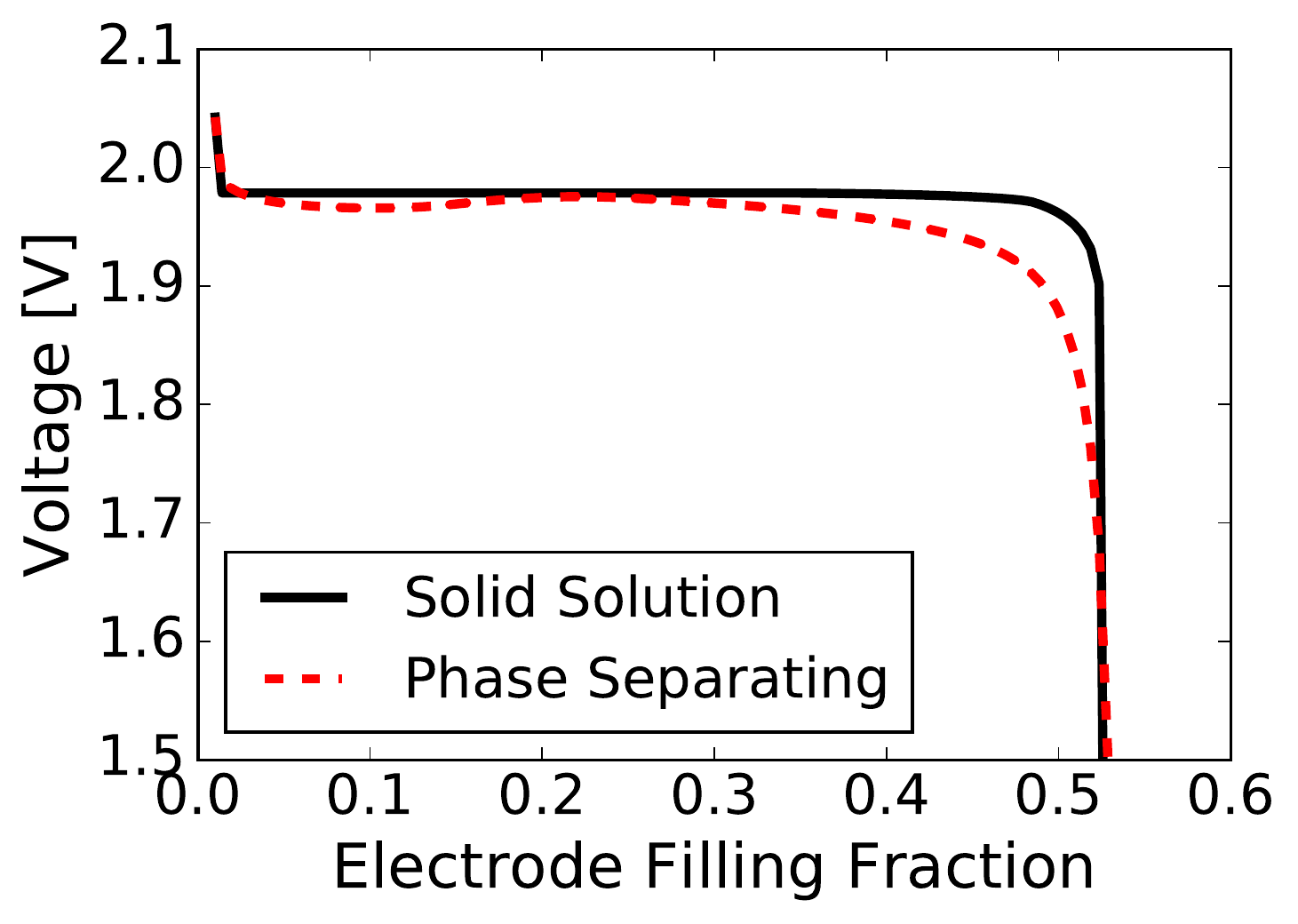}
    (b)
    \includegraphics[width=0.4\textwidth]{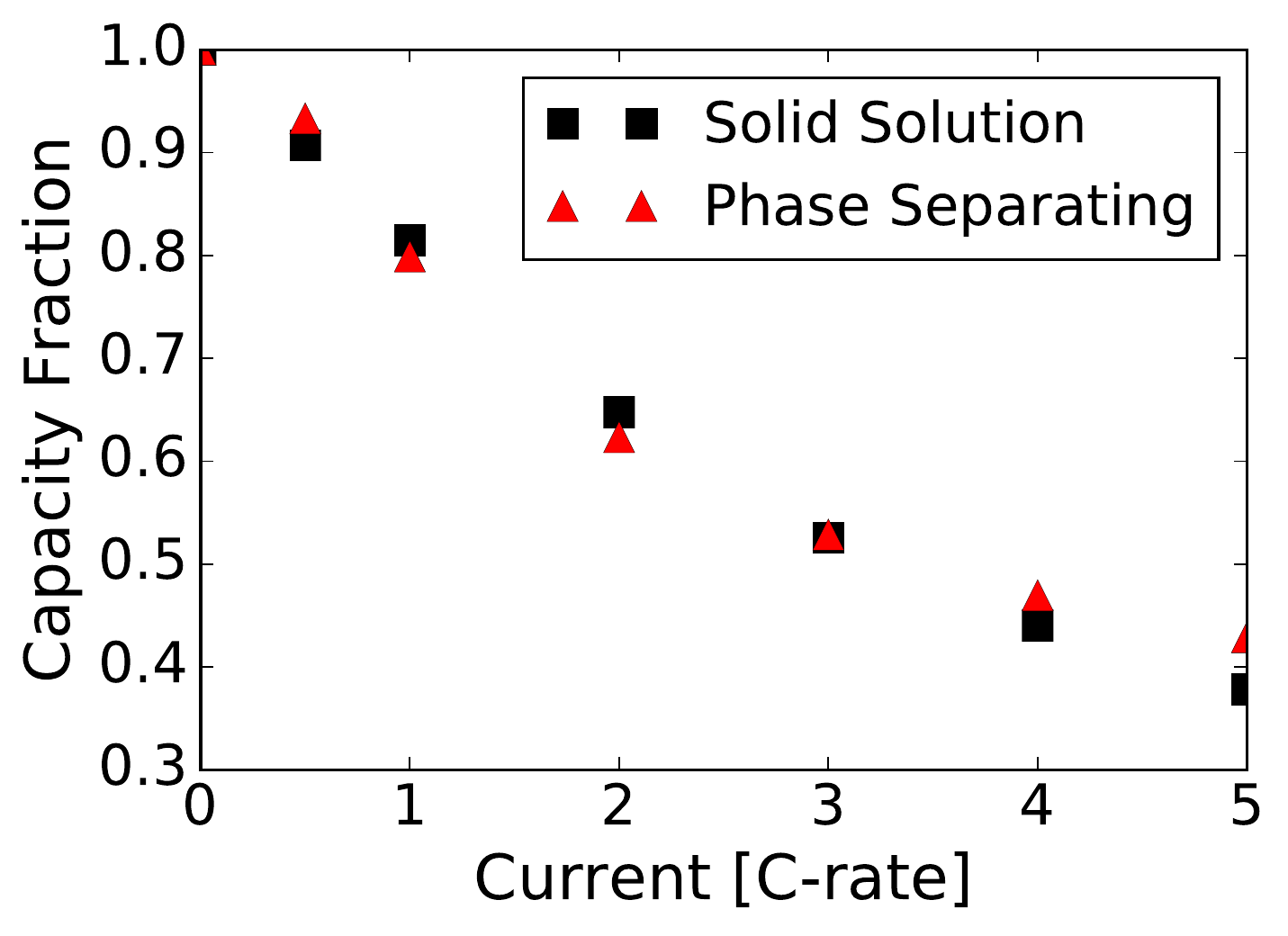}
    \caption{Discharge characteristics of solid solution and phase separating electrodes using constant exchange current density. Representative polarization curve at $3\ \mathrm{C}$ in (a) and dependence of capacity on current in (b).}
    \label{fig:ssps_v_3C}
\end{figure}

In order to minimize conflating factors from the reaction kinetics, we model both phase separating and solid solution particles using the symmetric ($\alpha = 0.5$) Butler-Volmer model with a constant exchange current density, $i_0 = 1\ \mathrm{A}/\mathrm{m}^2$, leading to minimal reaction losses in either system.
We will consider the case that intercalation occurs at the surface and phase separation occurs within the bulk, indicating it can be described by the Cahn-Hilliard reaction model~\cite{bazant2013,zeng2014}.
The transport of solid solution material can be described using Fickian diffusion.
For the solid solution, we use constant $D_\mathrm{chem}=1\times10^{-16}\ \mathrm{m}^2/\mathrm{s}$, and for the phase separating case, we use Eq.~\ref{eq:CHRflux_lattice}, which approximately matches the concentration-independence of the chemical diffusivity in the solid solution regimes~\cite{smith2017_intercalationdraft}.
We adjust $D_0$ until we get a similar dependence of available battery capacity as a function of current (Fig.~\ref{fig:ssps_v_3C} (b)), where capacity is defined as the electrode filling fraction when the voltage reaches a $1.5\ \mathrm{V}$ cutoff.
We find $D_0 = 8\times10^{-16}\ \mathrm{m}^2/\mathrm{s}$ reflecting the larger value of transport prefactor required to obtain similar fluxes in phase separating systems because of the smaller gradients which can arise in their end member phases compared to solid solution particles (Fig.~\ref{fig:ssps_csld_3C}).

We use a cell geometry with thin electrodes to minimize electrolyte limitations and assume no electron limitations in the porous electrode or reaction resistance on the lithium foil counter electrode.
The porous separator and cathode are $15\ \mu\mathrm{m}$ and $20\ \mu\mathrm{m}$ with porosities of $0.8$ and $0.2$ respectively.
The loading fraction in the cathode is taken to be 0.7 with the Bruggeman exponent $a=-0.5$.
We model the electrolyte using concentrated solution theory as described in Section~\ref{sec:model_elyte} using parameters as determined by Valo̸en and Reimers~\cite{valoen2005} but replacing the conductivity with that measured by Bernardi and Go~\cite{bernardi2011}.

\begin{figure}[h]
    \centering
    \includegraphics[width=0.4\textwidth]{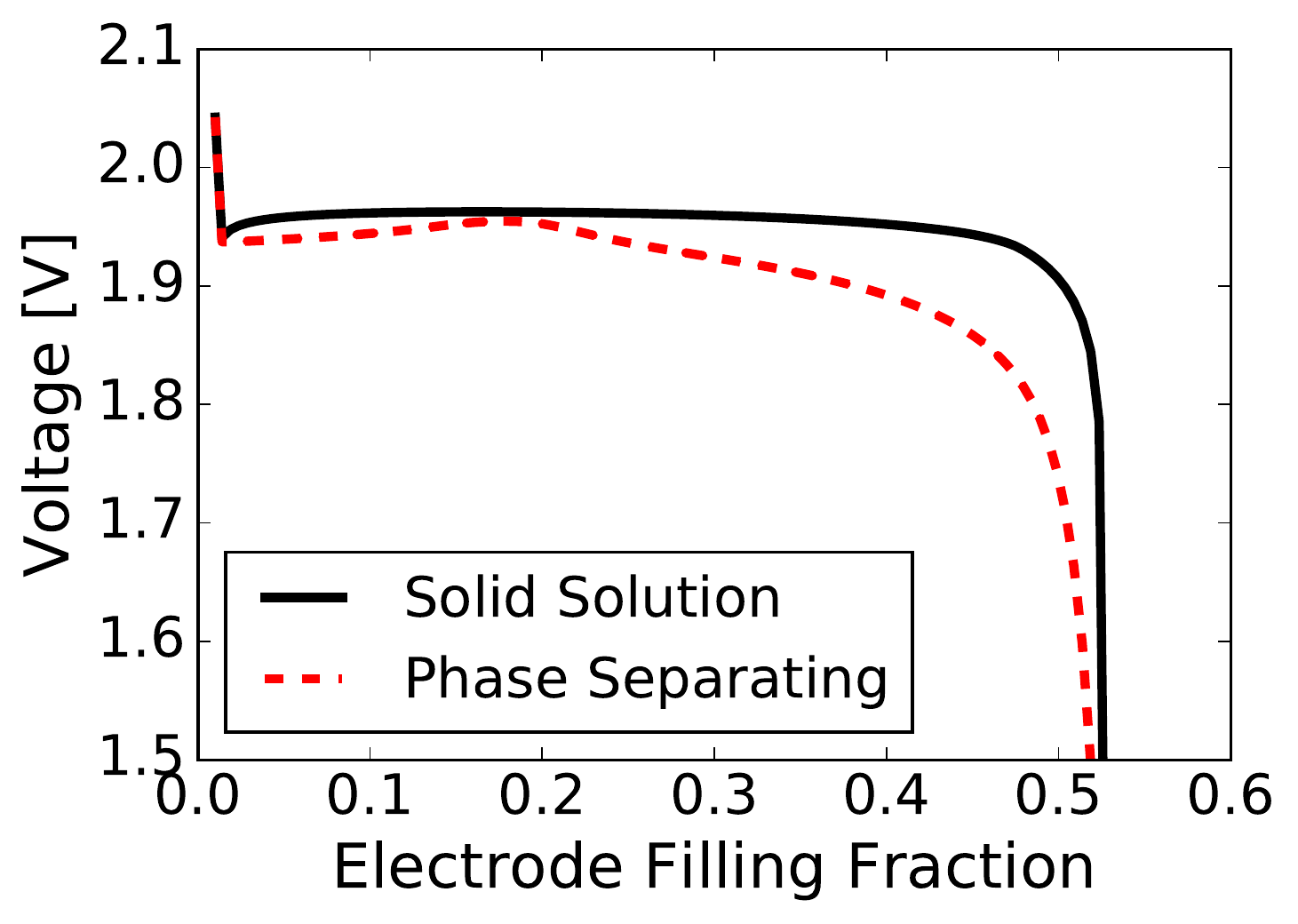}
    \caption{Discharge polarization curves at $3\ \mathrm{C}$ using concentration-dependent exchange current density, Eq~\ref{eq:ecd_conc}.}
%    \label{fig:capacity_crate}
    \label{fig:ssps_v_3C_bvmod01}
\end{figure}

Despite very different models for the transport in the solid phase, it is interesting to note that some macroscopic characteristics of the modeled cell are quite similar.
For example a representative cell voltage polarization curve and the dependence of cell capacity as a function of current show similar behavior (Figure~\ref{fig:ssps_v_3C}), indicating that macroscopic cell data alone would not differentiate between these models.
We find in a companion paper that this observation holds for similar models even in the case of non-negligible reaction resistances which are dominated by film resistance~\cite{thomas-alyea2017_submitted}.
Of note, in the companion paper, we used Eq.~\ref{eq:Ngradmu} with constant prefactor, i.e.\ $D_{0,i}\gamma_i\wt{c}_i/\gamma_{\ddagger,i}^d = \cnst$, which we found here unable to produce a dependence of capacity on current similar to the solid solution with constant chemical diffusivity.
Despite the similarity in Figure~\ref{fig:ssps_v_3C}, we note that the models give very different predictions of the behavior at the particle level, shown in Figure~\ref{fig:ssps_csld_3C} in which we show both a snapshot of a representative concentration profile for each of the simulations and also a representative profile of the evolution of surface concentration in one of the particles.
Because the surface concentration also affects other model behavior, including the reaction rate, we show in Figure~\ref{fig:ssps_v_3C_bvmod01} that when both particles are simulated using the concentration-dependent exchange current density in Eq.~\ref{eq:ecd_conc}, they diverge much more significantly in their predictions.
A consistent coupling to stresses and strains would further differentiate the models, as the concentration profile throughout the particle directly couples to the stress profile~\cite{garcia2004,bower2011finite,cogswell2012,dileo2014}.

\subsection{Concentrated and dilute solutions with phase separating materials}
\label{sec:cstdil}
Here, to focus on the electrolyte, we consider the same (phase separating) electrode materials in dilute and concentrated electrolyte solutions.
Dargaville and Farrell first considered CHR and ACR phase field particles in porous electrodes using Stefan-Maxwell electrolytes~\cite{dargaville2013comparison}, and we briefly compare the dilute approach previously used by Bazant and co-workers~\cite{ferguson2012,ferguson2014,li2014} with the concentrated model here.
We use the same phase separating material as that studied Section~\ref{sec:ssps} with constant exchange current density but simulate it in a battery with a longer electrode of $200\ \mu\mathrm{m}$.
The concentrated electrolyte model is identical to that used in Section~\ref{sec:ssps}, and the dilute model is analogous to that model but replacing the fit function of the electrolyte conductivity with the linear dependence predicted by dilute solutions~\cite{newman2004}, neglecting concentration dependence of the electrolyte transport coefficients, and using $D_{\ell,+} = 2.42\times10^{-10}\ \mathrm{m}^2/\mathrm{s}$ and $D_{\ell,-} = 3.95\times10^{-10}\ \mathrm{m}^2/\mathrm{s}$, which corresponds to the concentrated solution model with a typical value of $D_\ell = 3\times10^{-10}\ \mathrm{m}^2/\mathrm{s}$ and $t_+^0 = 0.38$.
\begin{figure}[h]
    \centering
    (a)
    \includegraphics[width=0.4\textwidth]{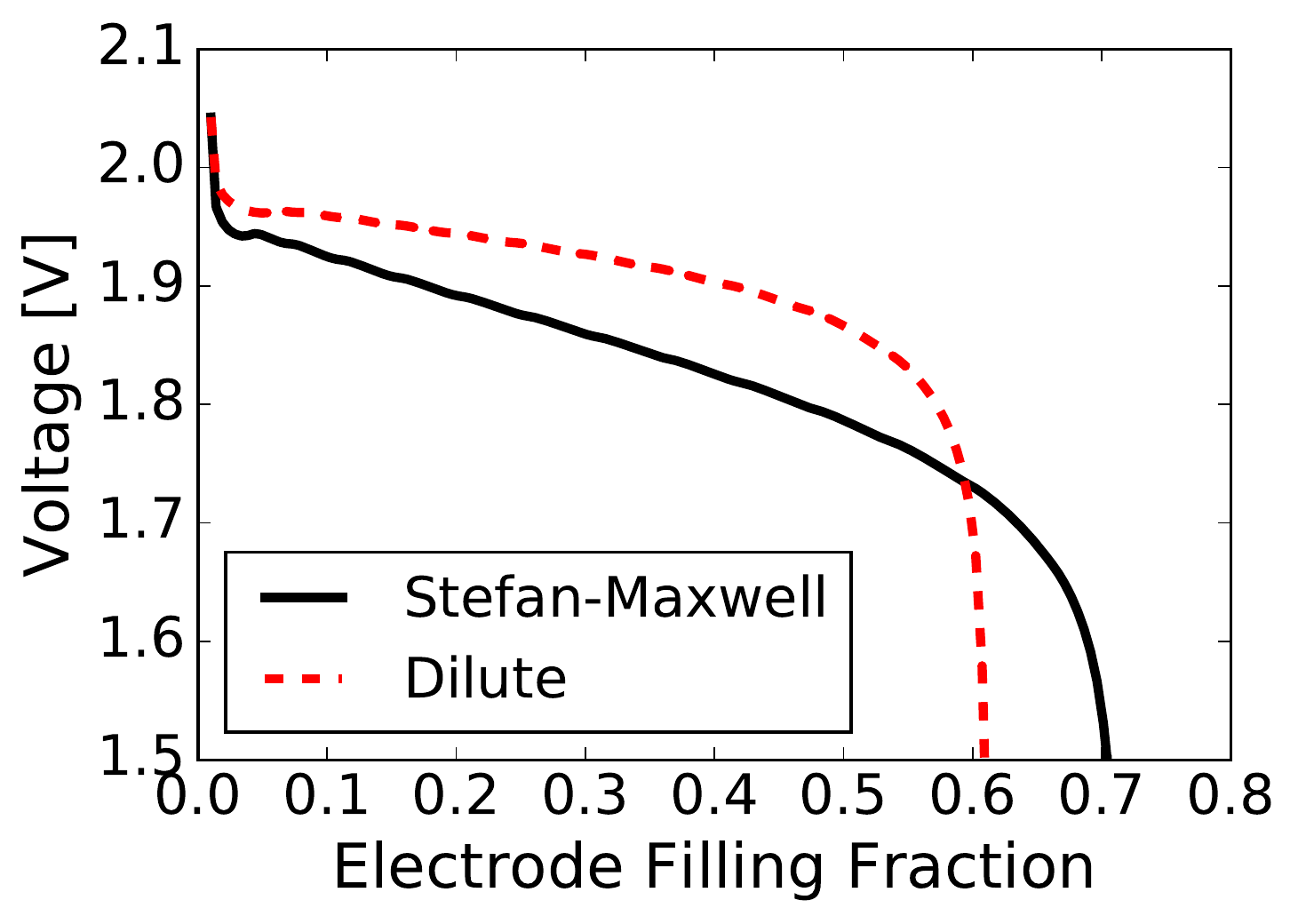}
    (b)
    \includegraphics[width=0.4\textwidth]{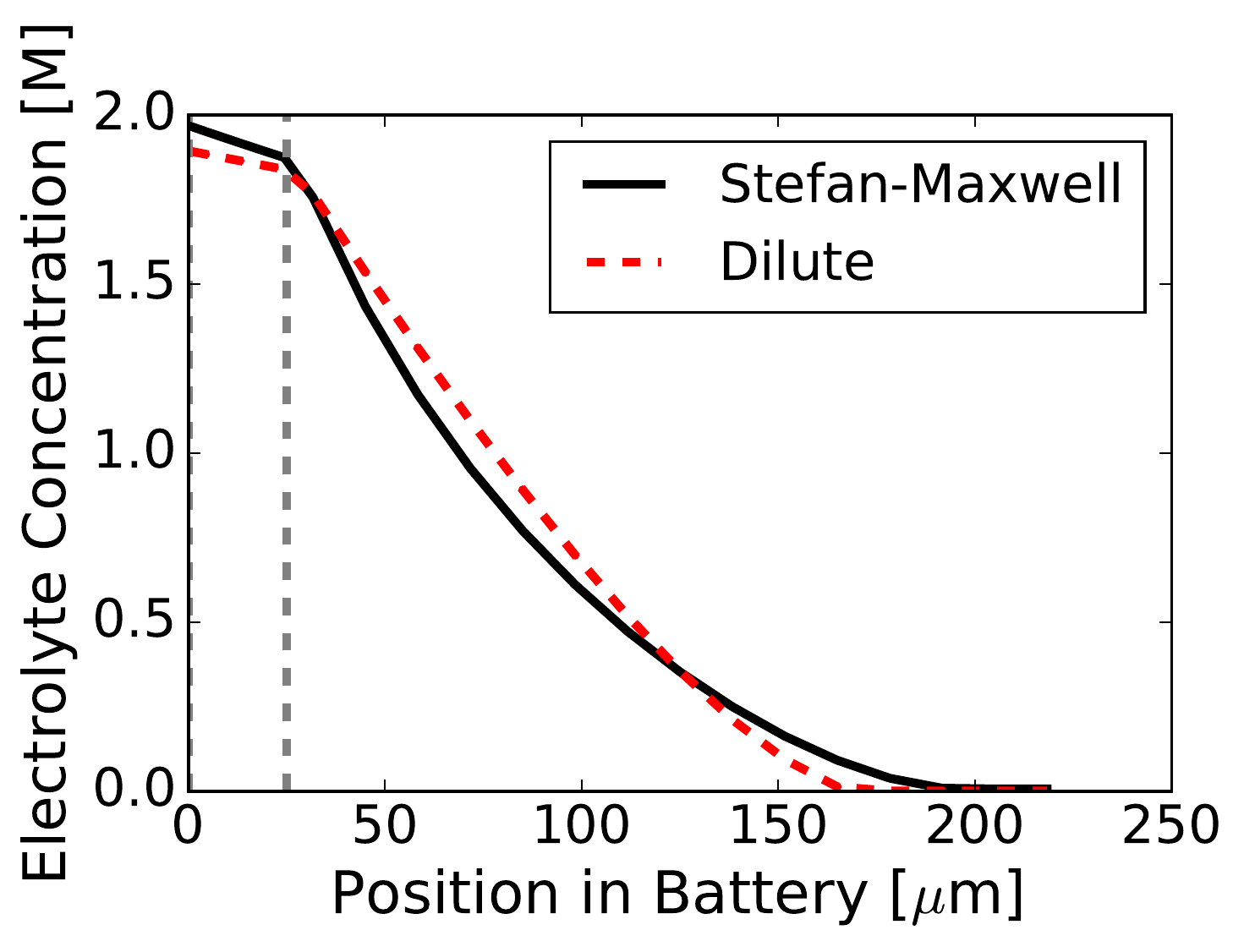}
    \caption{Comparison of phase separating porous electrodes using dilute and concentrated solution theory to model the electrolyte. Polarization curves (a) and final electrolyte concentration profiles (b) are presented for a $1\ \mathrm{C}$ discharge. In (b) the gray vertical dashed lines indicate the edge of the separator.}
    \label{fig:cstdil_v_cf}
\end{figure}

In this situation, the electrolyte transport is limiting, so we expect substantial differences between the two models, which we see in Figure~\ref{fig:cstdil_v_cf}.
Because the electrolyte transport is much more limiting than the solid phase transport, the results are qualitatively similar to those of Valo̸en and Reimers despite their use of a solid solution active material model instead of the phase separating model here~\cite{valoen2005}.
As in Figure~\ref{fig:cstdil_v_cf} (a), they found the concentrated model leads to a more gradual decrease in voltage over the cell discharge but a larger overall capacity.
This highlights the importance of using the more consistent concentrated solution theory approach in situations with substantial electrolyte polarization.

\subsection{Full Cell LiFePO$_4$/C}
\label{sec:lfp-c}
To demonstrate the versatility of the software, we present here the result of a battery simulation with two porous electrodes using two very different solid material models for each one.
In the cathode, we use an Allen-Cahn reaction model of LiFePO$_4$ (LFP) which was developed by Bazant and co-workers~\cite{burch2008,bai2011,cogswell2012,ferguson2014} and used to explain \emph{in situ} measurements of mosaic phase transformations in porous electrodes~\cite{li2014}.
In the anode, we simulate graphite using a 2-layer Cahn-Hilliard reaction model~\cite{ferguson2014,smith2017_intercalationdraft} as used to capture the experimental intercalation of a single crystal of graphite~\cite{guo2016}.
Although we have found a variation of this model to better describe the behavior of porous graphite electrodes~\cite{thomas-alyea2017_submitted}, we use it here to highlight the ease of implementing and using very different active material models within the same simulation using MPET\@.

We use the same concentrated electrolyte model as in the previous two sections.
The lengths of the anode, separator, and cathode are $100\ \mu\mathrm{m}$, $20\ \mu\mathrm{m}$, and $150\ \mu\mathrm{m}$.
The porosities are $0.15$, $0.4$, and $0.2$, and the loading percents of the anode and cathode are $0.9$ and $0.7$.
Because the solid phase models are described in detail in the previous work, we only briefly introduce them here.
The cathode particles are ACR particles described by a regular solution with an effective stress term, designed to approximately capture the tilting of the single-particle voltage plateau resulting from stresses~\cite{cogswell2012,cogswell2013,ferguson2014},
\begin{align}
    \mu_\mathrm{LFP} = k_\mathrm{B}T\ln\left( \frac{\wt{c}}{1-\wt{c}} \right) + \Omega\left( 1-2\wt{c} \right) + \frac{B}{c_\mathrm{\max,LFP}}\left( \wt{c} - \ovl{c} \right) - \frac{\kappa_\mathrm{LFP}}{c_\mathrm{\max,LFP}}\nabla^2\wt{c} + \mu^\Theta_\mathrm{LFP}
    \label{}
\end{align}
where $\Omega = 4.51k_\mathrm{B}T$, the stress coefficient $B = 0.19\ \mathrm{GPa}$, $c_\mathrm{\max,LFP} = 23\ \mathrm{M}$, $\ovl{c}$ is the average filling fraction in the particle, $\kappa_\mathrm{LFP} = 5\times10^{-10}\ \mathrm{J}/\mathrm{m}$, and $\mu^\Theta_\mathrm{LFP} = -3.4\ \mathrm{eV}$ with respect to Li/Li$^+$.
The anode particles are described by two-layer CHR transport with a two-parameter regular solution model including inter-layer repulsion terms to capture the staged structure found in intercalated graphite~\cite{dresselhaus1981}.
The flux within each layer is given by Eq.~\ref{eq:CHRflux_lattice}, and the diffusional chemical potential of each is given by
\begin{align}
    \mu_{G,i} = k_\mathrm{B}T\ln\left( \frac{\wt{c}_i}{1-\wt{c}_i} \right)
    + \Omega_a\left( 1-2\wt{c}_i \right) - \frac{2\kappa_\mathrm{G}}{c_\mathrm{\max,G}}\nabla^2\wt{c}_i
    + \Omega_b\wt{c}_j + \Omega_c\left( 1-2\wt{c}_i \right)\wt{c}_j\left( 1-\wt{c}_j \right)
    + \mu^\Theta_\mathrm{G}
    \label{}
\end{align}
where $j\ne i$ and $i \in \left\{ 1,2 \right\}$ and $\wt{c}_i$ represents the filling fraction of layer $i$.
The parameters for graphite come from refs.~\cite{ferguson2014,guo2016} and are $\Omega_a = 3.4k_\mathrm{B}T$, $\Omega_b = 1.4k_\mathrm{B}T$, $\Omega_c = 20k_\mathrm{B}T$, $c_\mathrm{\max,G} = 28.2\ \mathrm{M}$, $\kappa_\mathrm{G} = 4\times10^{-7}\ \mathrm{J}/\mathrm{m}$, and $\mu^\Theta_\mathrm{G} = -0.12\ \mathrm{eV}$ with respect to Li/Li$^+$.
In both materials, the reactions are governed by the symmetric ($\alpha=0.5$) Butler-Volmer, Eq.~\ref{eq:bv}, using Eq.~\ref{eq:ecd_act} for the exchange current density.
For LFP, we set $k_0 = 0.16\ \mathrm{A}/\mathrm{m}^2$ and use $\gamma_\ddagger$ = $1/\left( 1-\wt{c} \right)$ as proposed by Bai et al.~\cite{bai2011}.
For graphite, we set $k_0=10\ \mathrm{A}/\mathrm{m}^2$ and use $\gamma_{\ddagger,i} = 1/\left( \wt{c}_i\left( 1-\wt{c}_i \right) \right)$ as we initially used in Guo et al.~\cite{guo2016} and reexamined recently to suggest alternate models for practical battery simulation~\cite{smith2017_intercalationdraft}.
\begin{figure}[h]
    \centering
    \includegraphics[width=0.4\textwidth]{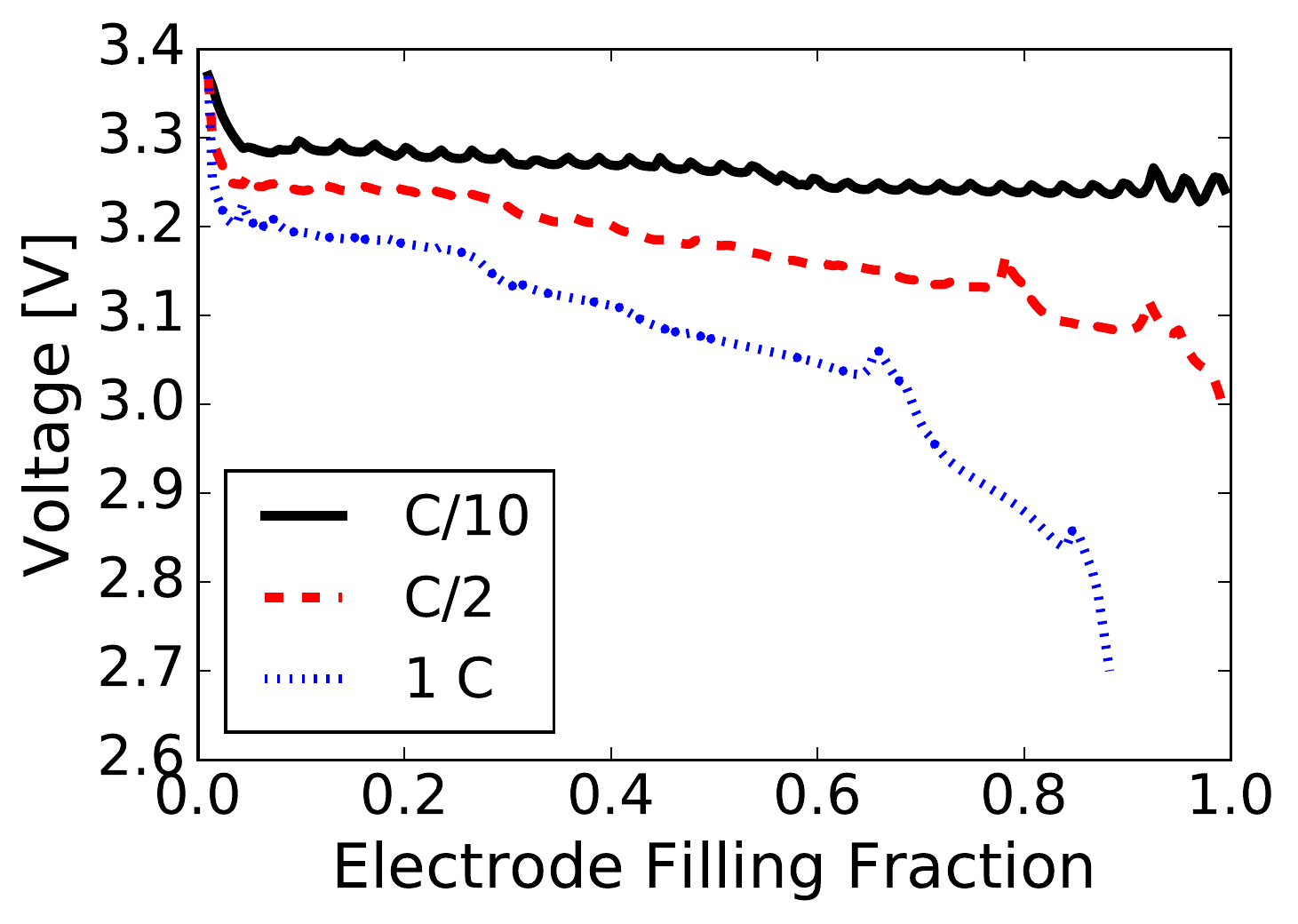}
    \caption{Simulated Graphite-LiFePO$_4$ cell discharge polarization curves. The bumps in the $\mathrm{C}/10$ curve arise from discrete particle filling events in the cathode, and the bumps in the $\mathrm{C}/2$ and $1\ \mathrm{C}$ curves arise from the graphite reaction activity coefficient model~\cite{smith2017_intercalationdraft}. The electrode filling fraction refers to the limiting electrode, the cathode.}
    \label{fig:lfp-c-pol}
\end{figure}
\begin{figure}[h]
    \centering
    (a)
    \includegraphics[width=0.7\textwidth]{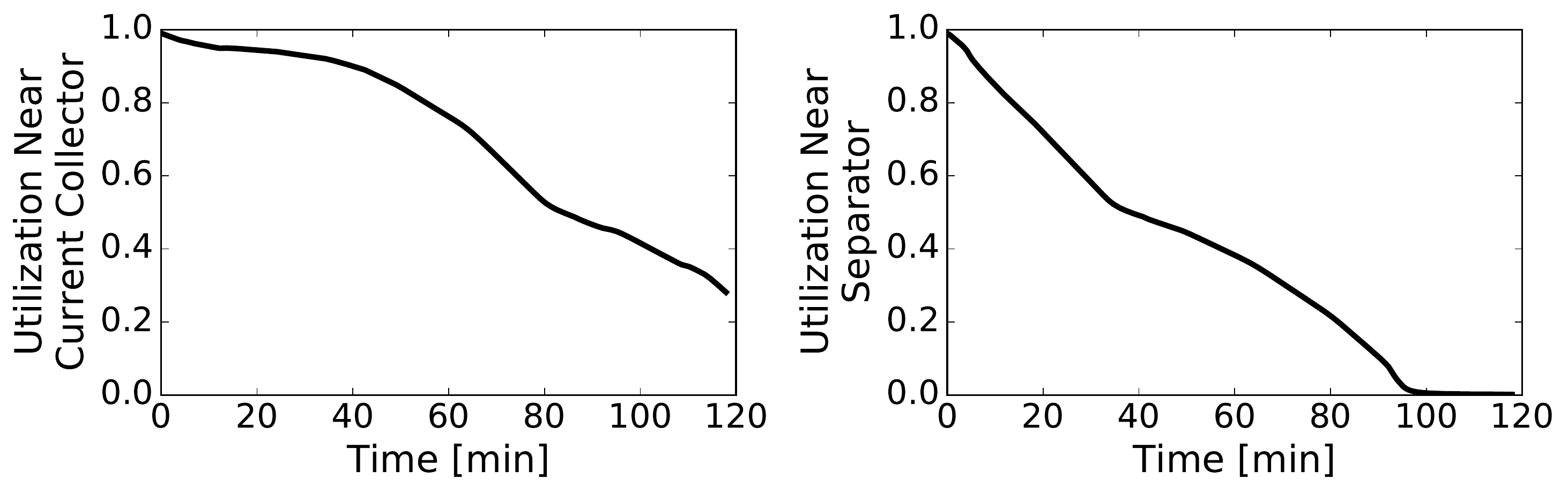}
    \\
    (b)
    \includegraphics[width=0.7\textwidth]{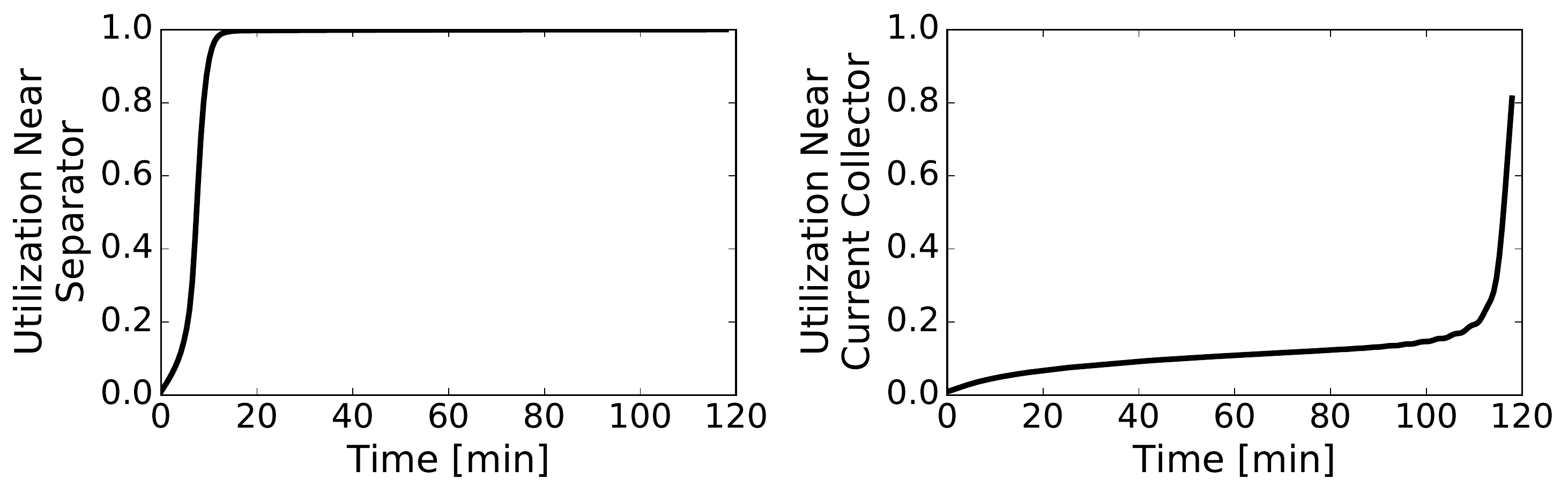}
    \caption{Utilization (average filling fraction) of particles on both ends of each electrode at $\mathrm{C}/2$. The anode is depicted on top (a) and the cathode on bottom (b).}
    \label{fig:lfp-c-util}
\end{figure}

We present overall cell polarization curves for a selection of currents in Figure~\ref{fig:lfp-c-pol}.
The thick electrodes and low porosity make the cell experience strong electrolyte polarization at the moderate simulated currents, causing most of the losses in the cell.
The cell capacity is limited by the cathode, evidenced by the graphite-caused shift in the $\mathrm{C}/10$ voltage profile occurring at a cathode filling fraction larger than $0.5$.
In addition, we see erratic profiles at all simulated rates.
At the lowest rates, this is primarily related to the mosaic transformation of the LFP particles along the length of the electrolyte, with one representative particle typically receiving almost all the current in the electrode, as seen experimentally~\cite{chueh2013,li2014} and explained theoretically~\cite{dreyer2010,ferguson2014}.
This behavior can be seen in Figure~\ref{fig:lfp-c-util} (b) in which the cathode particles show sharp filling processes related only to their position in the electrode, because they do not have size variation as used in Ferguson and Bazant~\cite{ferguson2014}.
In contrast, the anode particles behave more like the CHR particles studied in the sections above, with more gradual filling processes governed more weakly by their position in the electrode.
At higher rates, the cathode still demonstrates a mosaic transformation, but the voltage curves smooth out because other losses dominate.
The voltage spikes result from the form of the reaction resistance causing short sections of low resistance.
This rate expression helped capture single particle experiments~\cite{guo2016} but is unable to capture experiments on a full porous electrode, where we also used a simplified version of the graphite model~\cite{thomas-alyea2017_submitted}.

\subsection{Electrochemical reactions}
\label{sec:bv-mhc}
Battery models are typically constructed assuming Butler-Volmer reaction kinetics~\cite{botte2000mathematical,thomas2002}, although recent evidence suggests Marcus kinetics may better describe some electrochemical reactions~\cite{chidsey1991,bai2014}.
One possible reason for the use of the Butler-Volmer form is that the models do not deviate until moderate to large reaction driving forces, depending on the reaction~\cite{henstridge2012marcus}.
A second is that the expression for calculating Marcus style kinetics at an electrode involves an improper integral~\cite{chidsey1991}, which is cumbersome to evaluate, especially within porous electrode simulations in which reaction rates are evaluated many times over the course of a simulation.
Nevertheless, this Marcus-Hush-Chidsey (MHC) reaction model has been approximated with a mathematically simple and accurate expression~\cite{zeng2014simple}, enabling its practical use in porous electrode battery models, which we demonstrate here.
The primary effect of MHC kinetics is a downward curvature instead of the linear Tafel slope associated with Butler-Volmer reaction kinetics.

\begin{figure}[h]
    \centering
    \includegraphics[width=0.4\textwidth]{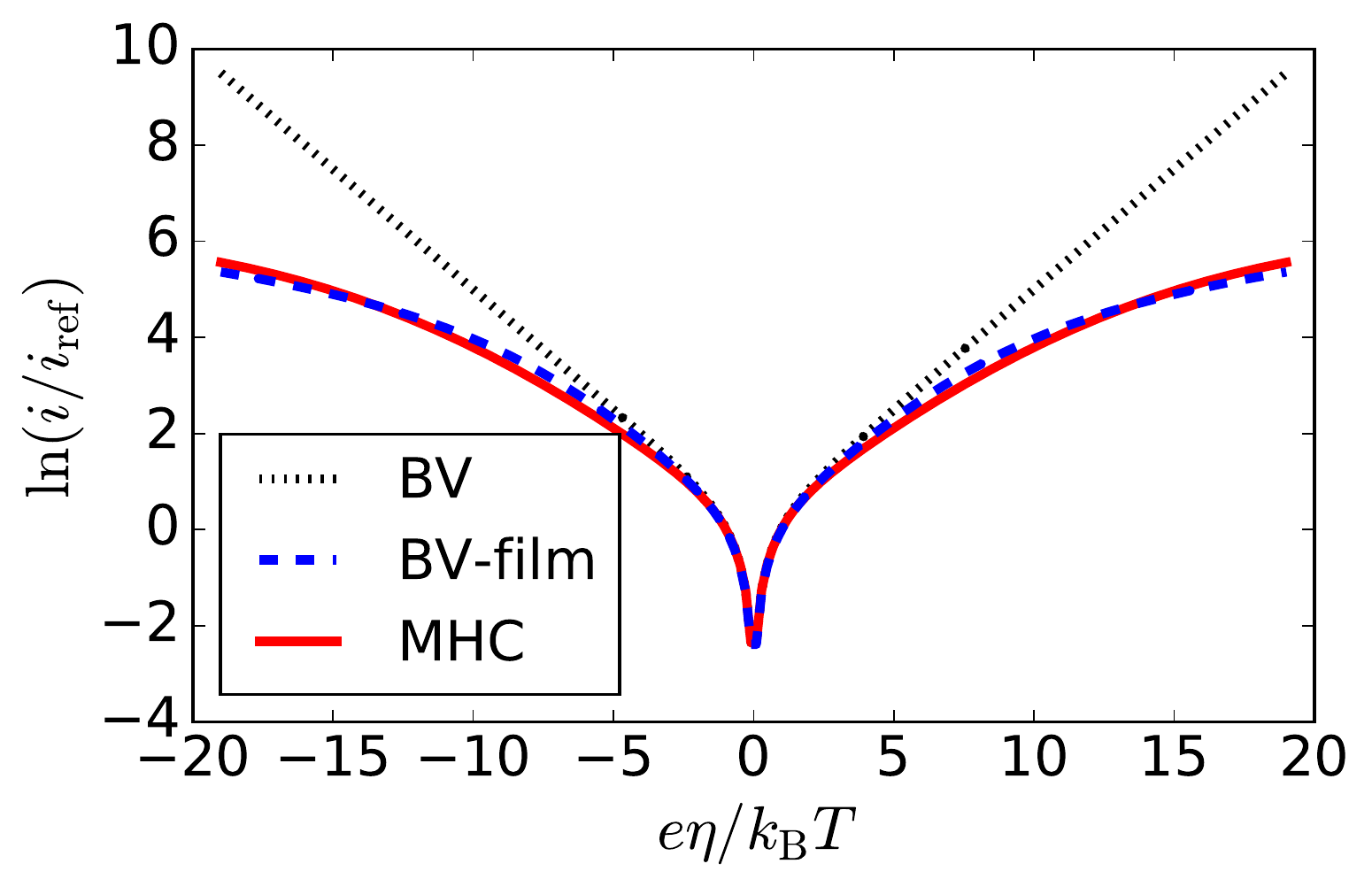}
    \caption{Comparison of reaction models.}
    \label{fig:rxns}
\end{figure}

In order to accurately capture electrochemical data, the Butler-Volmer expression is commonly modified with a film resistance as in Eq.~\ref{eq:Rfilm} to capture a series resistance associated with any film at the surface.
Curiously, this modified expression could be interpreted as a way to introduce curvature to the Tafel plot in a way that can look similar to MHC kinetics over a range of driving forces.
For example, in Figure~\ref{fig:rxns}, we compare Butler-Volmer with and without film resistance (BV and BV-film respectively) with the MHC expression, all using the same exchange current density and at fixed concentrations.
In the figure, we use constant reaction rate prefactors, $i_0 = i_M = 1\ \mathrm{A/m}^2$ and $R_\mathrm{film} = 0.001\ \Omega\ \mathrm{m}^2$, a value similar to that used to match electrochemical data using porous electrode modeling~\cite{thomas2003thermal,srinivasan2004,albertus2009}.
The MHC kinetics use the same exchange current density and a reorganization energy, $\lambda = 18k_\mathrm{B}T$, between that approximated for LiFePO$_4$~\cite{bai2014} and values used in other interfacial reactions~\cite{chidsey1991,henstridge2012marcus}.
The limiting behaviors for the film resistance and MHC models differ substantially, as the film model approaches a linear resistor at high driving forces, whereas the MHC current saturates at a constant value.
However, over a relatively broad range of driving forces, the two predict similar currents when neglecting concentration effects, which suggests that some variant of the MHC model may be able to capture some electrochemical data using a microscopically derived model.
We should note that the concentration dependence of the MHC model differs from that of the Butler-Volmer model with film resistance, so comparisons are not completely straightforward.
Nevertheless, use of the MHC expression has the advantage that it also makes clear connections to surface properties~\cite{kuznetsov_book} which could be engineered, suggesting possible improvement paths for materials with slow kinetics.

To illustrate the application of MHC kinetics in a porous electrode, we simulate a current-pulse discharge process of the idealized phase separating material studied in Section~\ref{sec:ssps} using both the Butler-Volmer and MHC models as well as the Butler-Volmer model with a film resistance.
As mentioned above, the concentration dependence of the MHC model is non-trivial, as the departure from the formal potential, $\eta_f$, is offset from the activation overpotential.
In addition, the prefactor, $i_M = k_M/\gamma_{M,\ddagger}$, may also have concentration dependence related to concentrated solution effects captured in $\gamma_{M,\ddagger}$.
In order to study the two with the least impact from the different concentration dependence, we choose to modify $i_M$ here to match the exchange current densities of the Butler-Volmer and MHC models.
This could be interpreted as choosing a particular model for $\gamma_{M,\ddagger}$, but we refrain from assigning physical meaning to the choice used here and make the selection to enable the simplest comparison between the two models.

\begin{figure}[h]
    \centering
    (a)
    \includegraphics[width=0.4\textwidth]{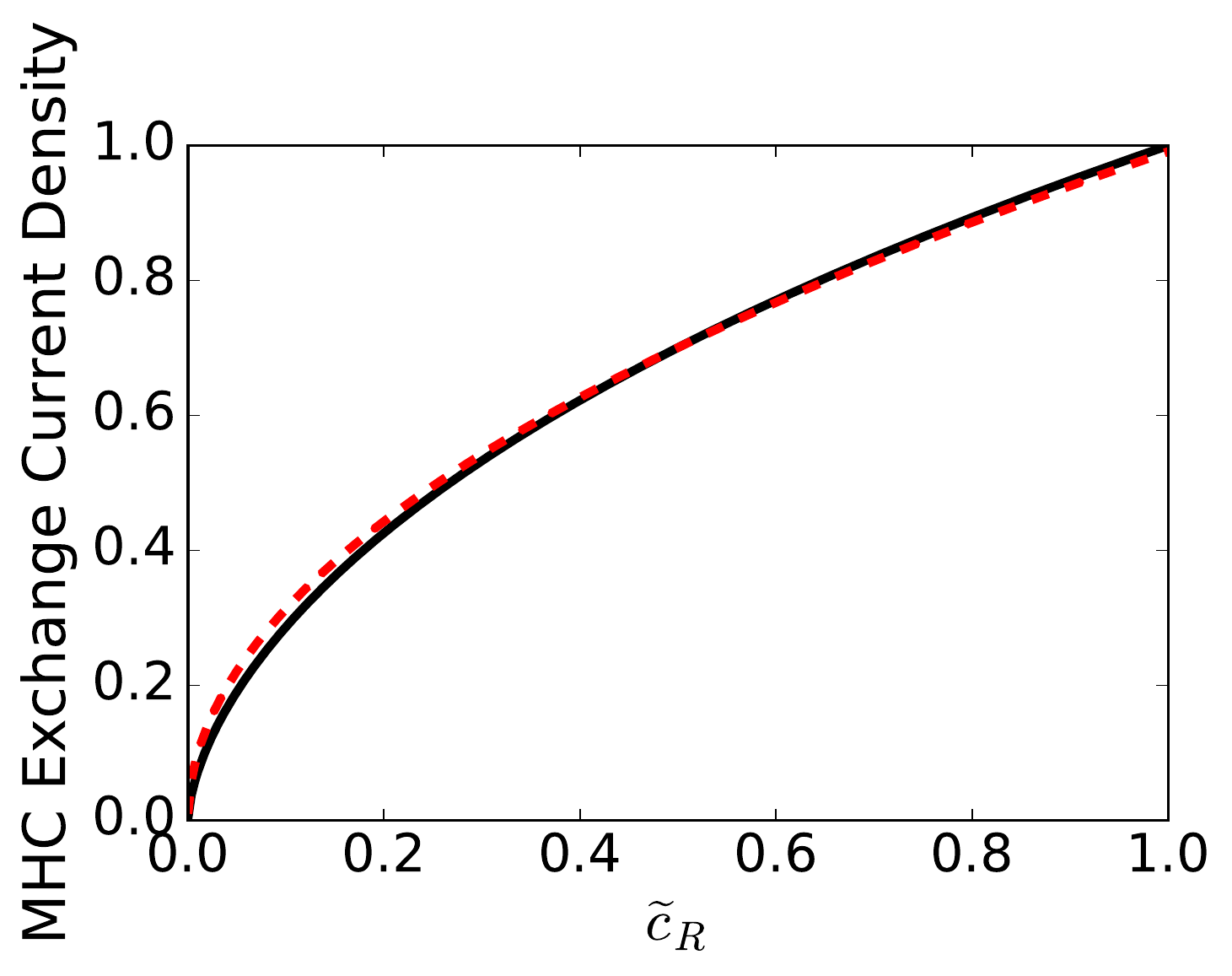}
    (b)
    \includegraphics[width=0.4\textwidth]{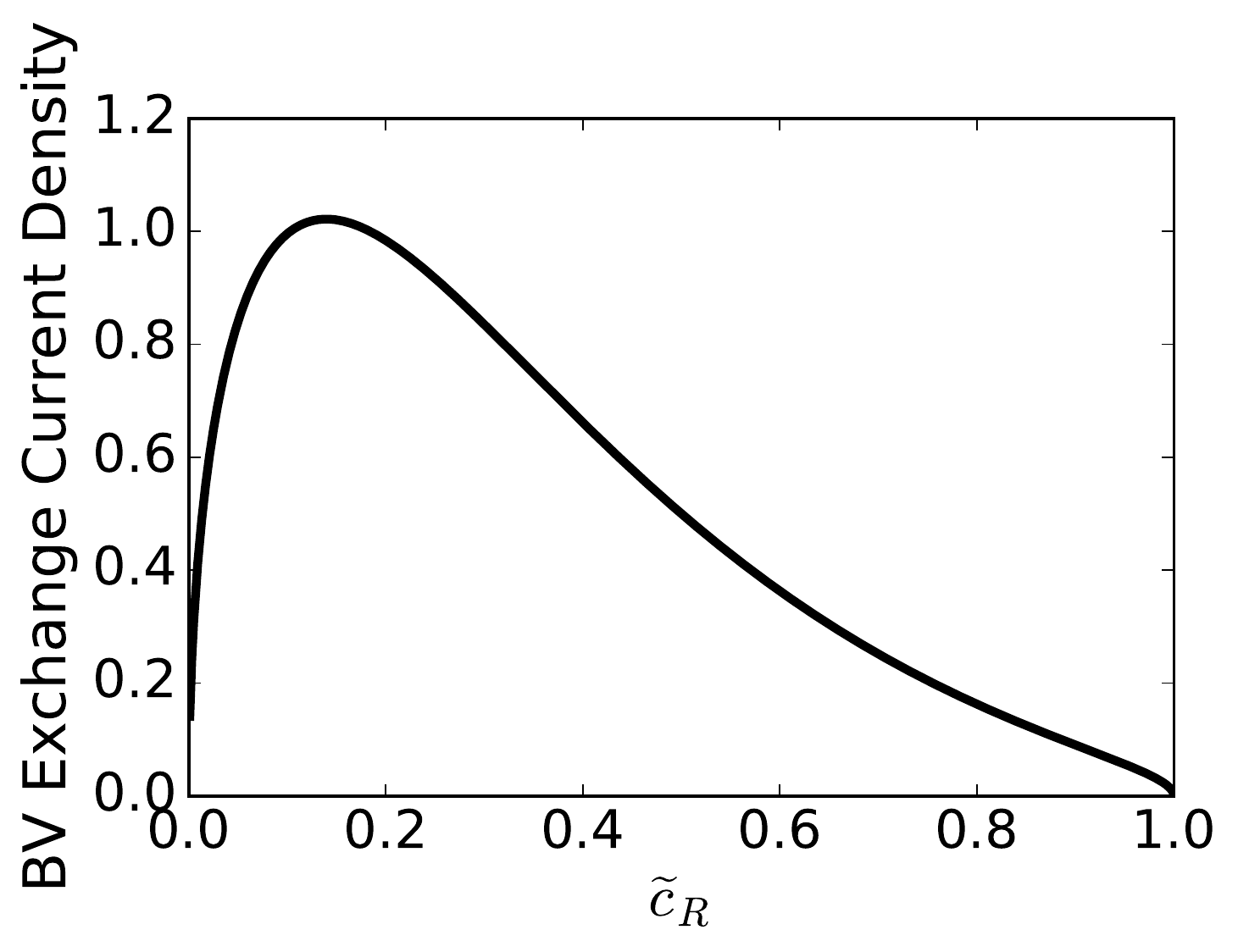}
    \caption{Exchange current densities as a function of concentration of the reduced species (e.g.\ intercalated lithium). In (a) the MHC exchange current density using constant prefactor, $i_M$, is normalized to its value at maximum solid filling, and the red dashed line is a fitted square root function. In (b) the Butler-Volmer exchange current density is scaled to the reaction rate prefactor, $k_0$, and assumes negligible contribution from concentration gradients.}
    \label{fig:mhc_ecd}
\end{figure}

First, we plot the exchange current density of the MHC model using constant $i_M$ in Figure~\ref{fig:mhc_ecd}~(a) and note that it has a dependence on the reduced species (intercalated lithium) that is approximately a square root.
This applies to both $\wt{c}_R$ and $\wt{c}_O$.
Thus, in order to approximately specify an arbitrary exchange current density for the MHC model, we can simply divide a chosen function by $\sqrt{\wt{c}_R\wt{c}_O}$ and scale the magnitude accordingly.
To connect to our previous work, we choose to compare the (symmetric) Butler-Volmer form based on species activities, Eq.~\ref{eq:bv} with Eq.~\ref{eq:ecd_act}.
Thus, for the MHC model, we choose
\begin{align}
    i_M = \frac{k_{M}n{\left( a_{O}a_\mathrm{e}^n \right)}^{1-\alpha}a_{R}^\alpha}{\gamma_{\ddagger}\sqrt{\wt{c}_R\wt{c}_O}},
\end{align}
with $n=1$ and $\alpha=0.5$,
which very nearly matches the concentration and activity dependence of the exchange current densities of the two models, leaving the primary difference in the dependence on the activation overpotential, $\eta$.
We use $\gamma_\ddagger = 1/\left( 1-\wt{c}_R \right)$ following previous work on LFP~\cite{bai2011,cogswell2012} and supported by recent experiments~\cite{lim2016origin}.
This gives an exchange current density with broken symmetry around half filling, as depicted in Figure~\ref{fig:mhc_ecd}~(b).
%Thus, in order to approximately match the concentration dependence, we use a symmetric Butler-Volmer with exchange current density defined by Eq.~\ref{eq:ecd_conc}, and set $i_M = k_M\sqrt{c_\ell\left( 1-\wt{c} \right)}$ which very nearly matches the concentration dependence of the exchange current densities of the two models, leaving the primary difference in the dependence on the activation overpotential, $\eta$.
We choose a reaction reorganization energy for the MHC model $\lambda = 18k_\mathrm{B}T$ as in Figure~\ref{fig:rxns}.

In order to minimize the effect of solid phase transport losses, we modify the phase separating particles from Section~\ref{sec:ssps} to have $D_0 = 1\times10^{-14}\ \mathrm{m}^2/\mathrm{s}$.
We simulate a half-cell using the concentrated Stefan-Maxwell electrolyte model and with a lithium foil counter-electrode with fast kinetics.
Although fast lithium foil kinetics may be a poor assumption because of the small relative surface area, under constant current conditions, it would only add a current-dependent, constant additional voltage drop to both simulated systems.
The simulated cell has a separator of length $20\ \mu\mathrm{m}$ with porosity $0.8$ and a thin cathode of length $25\ \mu\mathrm{m}$ with porosity $0.2$ to minimize electrolyte transport losses and focus on the effect of the reaction kinetics.
The cathode loading percent, as above, is set to $0.7$, and we simulate uniform particles of radius $1\ \mu\mathrm{m}$ with $k_0 = 1\ \mathrm{A}/\mathrm{m}^2$ and $k_M$ chosen such that the magnitudes of the exchange current densities match.
We adjust the value of the film resistance from that used in Figure~\ref{fig:rxns} to $R_\mathrm{film} = 0.02\ \Omega\ \mathrm{m}^2$ to achieve a closer match between the MHC and Butler-Volmer model with the film while using the concentration-dependent exchange current densities neglected in Figure~\ref{fig:rxns}.
This value is higher than is commonly used and closer to that used in the introduction of the film resistance to battery modeling by Doyle et al.~\cite{doyle1996}.

\begin{figure}[h]
    \centering
    (a)
    \includegraphics[width=0.4\textwidth]{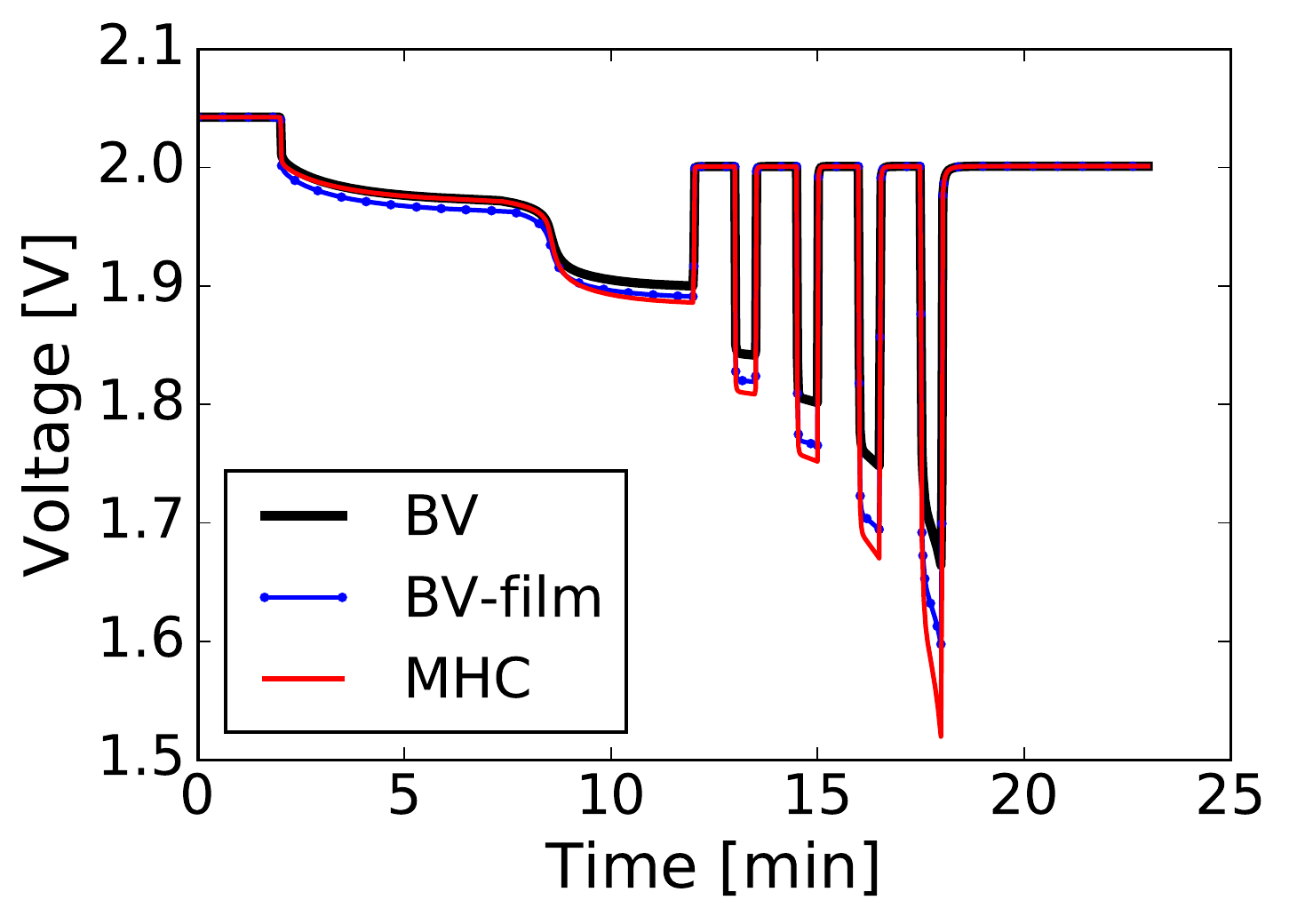}
    \\
    (b)
    \hspace{0.34in}\includegraphics[width=0.424\textwidth]{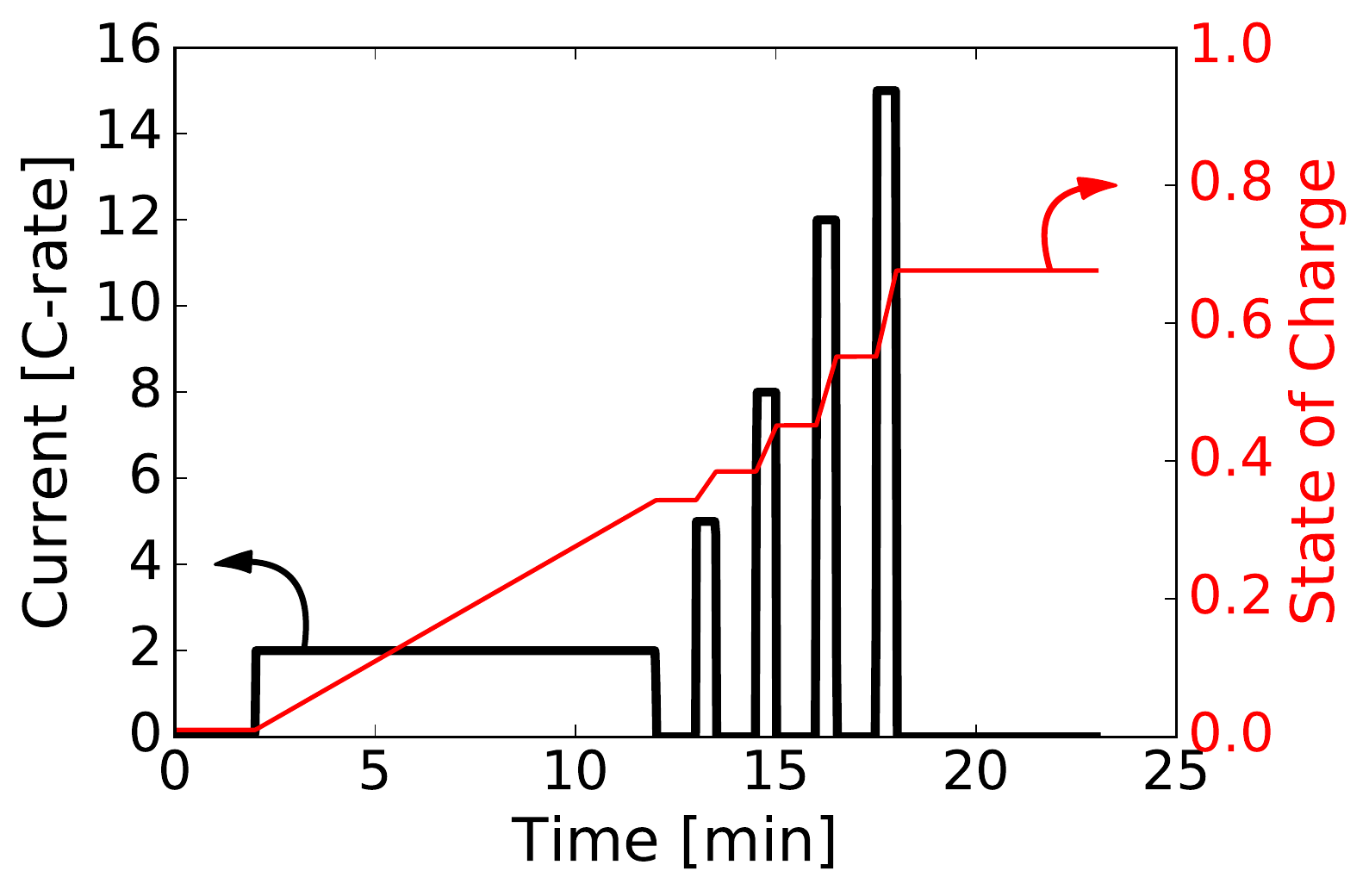}
    \caption{Comparison of reaction models within a porous electrode model. In (a) the system voltage response is plotted, while in (b) the specified input current and associated state of charge profiles are shown.}
    \label{fig:bv-mhc_vt}
\end{figure}

We expose the cell to the current profile and corresponding state of charge profile in Figure~\ref{fig:bv-mhc_vt} (b), and the output voltage profiles are shown in Figure~\ref{fig:bv-mhc_vt} (a).
Although the cell experiences moderate electrolyte polarization at the highest currents, those losses are similar for both models and do not explain the differences in the overall output voltage.
The initial current pulse at $2\ \mathrm{C}$ brings all the electrode particles to a state of charge within the miscibility gap, so the equilibrium voltage is given by the value of $-\mu^\Theta/e = 2\ \mathrm{V}$ which we see the cell relax to after the pulse.
We also can see the overlap of the BV and MHC profiles in the initial region of this pulse, where the surface concentrations are changing most substantially, confirming the match of both the magnitude and concentration dependence of the two reaction models in this lower-overpotential range when the two reaction models should overlap.
However, once the particles begin to phase separate at approximately $7.5\ \mathrm{min}$, the surface concentrations rise sharply, and the exchange current density decreases (Figure~\ref{fig:mhc_ecd}~(b)).
This causes increased reaction resistance in both models and an associated increase in required driving force, but the MHC model requires a larger increase in driving force, causing the initial departure of the models near $7.5\ \mathrm{min}$.
When the higher current pulses begin, we see further departure of the two reaction models with the MHC model showing substantially higher reaction losses in Figure~\ref{fig:bv-mhc_vt} (a) than the Butler-Volmer model.
The Butler-Volmer model with the film resistance leads to behavior between the two, predicting more resistance than the other models at low rates and intermediate resistance at higher rates, indicating that the model results depart enough to clearly distinguish in these conditions.
Still, the effect of the film resistance and MHC kinetics are qualitatively similar, suggesting that some of the system behavior which has been attributed to reaction film resistance may instead be caused by fundamental departures from Butler-Volmer reaction kinetics.

In the limit of $e\eta \ll \lambda$, the Butler-Volmer and MHC models predict identical results.
The value of $\lambda$ we used here is between a value calculated for LFP~\cite{bai2014} and values commonly found in calculations or fits to data for other systems~\cite{chidsey1991,henstridge2012marcus}.
Using this intermediate value and an MHC prefactor adjusted to give a more similar concentration dependence to Butler-Volmer expressions, we can see that the two models give noticeably different predictions on the simulated macroscopic system output.
This further highlights the importance of continuing the study of the two models in practical battery models, and the MPET software can help facilitate this.

\section{Summary and Conclusions}
\label{sec:concl}
Volume averaged porous electrode simulations provide insightful and industrially relevant ways to characterize battery behavior.
Despite shortcomings associated with information loss from the volume averaging process, the simplicity of the approach and associated speed of model development, implementation, and simulation run time motivate its continued use.
In this work, we have presented an open-source software package called MPET (Multiphase Porous Electrode Theory), which builds on foundations laid by John Newman and many others by describing the active materials with variational nonequilibrium thermodynamics~\cite{bazant2013,bazant2017thermodynamic} applied to porous electrodes~\cite{ferguson2012,ferguson2014}.
Despite the prevalence of this modeling approach, few open source options are available for simulating the model, particularly ones that are easy to modify with new thermodynamic models based on the powerful phase field formalism~\cite{chen2002phase}, adapted for electrochemical systems~\cite{bazant2013}.
With MPET, we aimed to address this gap by providing a software platform implementing nonequilibrium thermodynamics of porous electrodes with an open source code, written with a modular design to encourage use, modifications, and improvements.

Through a variety of examples, we have demonstrated some of key features of the MPET software.
First, we compared solid solution and phase field approaches to modeling active materials and demonstrated that the models can give similar macroscopic outputs under some situations, but deviate at the particle scale.
This leads to different predictions when the surface concentrations strongly affect reaction behavior.
Second, we reexamined the comparison of Stefan-Maxwell concentrated solution theory and dilute solution models of electrolyte transport in the context of electrodes made of simple phase separating particles.
Under strong electrolyte limitation, the difference in the model predictions is similar to that found in electrodes with solid solution particles.
Third, to highlight the ability of the software to describe unique and distinct solid models, we implemented a full cell simulation using models of graphite and iron phosphate presented in previous works.
Finally, we considered the effect of changing the reaction model from the typically used Butler-Volmer to Marcus-Hush-Chidsey (MHC) reaction kinetics, based on microscopic theories of electron transfer.
We demonstrated that, for some reasonably typical parameter values, the MHC reaction kinetics can look similar to Butler-Volmer reactions with a film resistance and can lead to substantial differences from the Butler-Volmer model in predicted battery behavior at high rates.

Natural extensions of this work involve implementing some of the features that other software options have and which researchers have found helpful in explaining data or better describing real systems.
For example, thermal effects can substantially affect cell behavior~\cite{pals1995thermal,gu2000thermal,thomas2003thermal,lai2011,prada2012,northrop2015,latz2015thermal,torchio2016lionsimba} and exploration of their coupling with Marcus kinetics would be interesting.
More complete thermal descriptions rely on temperature profiles over more than an individual cell layer~\cite{forgez2010thermal,pesaran2002battery}, so isothermal but non-constant temperature dependence would be a starting point.
Addition of material models properly coupling the stresses to the concentration profile would also be an opportunity to study the effects of electro-mechanical models with phase separation~\cite{cogswell2012,dileo2014} in porous electrodes.
Other capabilities, such as simulating electrochemical impedance outputs or others of the many additions that have been made to the original model implemented by Doyle et al.~\cite{doyle1993} could also be added.
For capacitive energy storage and desalination, the electrolyte model could also be extended to allow for diffuse charge in the electrode/electrolyte interface~\cite{biesheuvel2011diffuse,biesheuvel2012electrochemistry} or the double layers of charged porous separators~\cite{dydek2011,dydek2013,schmuck2015homogenization,yaroshchuk2012over}, which also activates additional mechanisms for ion transport by surface conduction and electro-osmotic flow~\cite{dydek2011,deng2013,mirzadeh2014enhanced}, which are neglected in traditional battery models.

In summary, MPET provides some new capabilities for battery simulation focused on recent developments in the modeling of active materials based on nonequilibrium thermodynamics.
It can also serve as a starting point for other researchers beginning to study in the area to make their own modifications and investigations.
By highlighting its capabilities, we have shown the value of a flexible simulation package to expand on the existing porous electrode theory and begun to examine the impact of those developments.

\section{Acknowledgments}
The research was supported by the Samsung-MIT Program for Materials Design in Energy Applications, and in part by the D3BATT program of the Toyota Research Institute. We thank E. Khoo and K. Conforti for proofreading the manuscript.

%\appendix
%
%\section{Extra content?}
%Maybe?

%\bibliographystyle{plain}
\bibliography{sources}
\end{document}